\documentclass[smallextended]{svjour3}

\usepackage{lmodern}

\usepackage[hidelinks]{hyperref}

\PassOptionsToPackage{intlimits}{mathtools}
\usepackage{mathtools}

\usepackage{amssymb}

\usepackage{bm}

\usepackage{graphicx}

\usepackage{csquotes}

\everymath{\displaystyle}

\newcommand\p[1]{{\left({#1}\right)}}

\newcommand\e{\mathrm{e}}
\newcommand\dif[1]{\mathrm{d}{#1}\mathop{}}

\newcommand\intl[3]{\int_{#2}^{#3}\dif{#1}}

\newcommand\E[1]{{\mathbb{E}\left[#1\right]}}

\newcommand\Prob[1]{{\mathbb{P}\left[#1\right]}}

\renewcommand\O[1]{\mathcal{O}\p{{#1}}}

\newcommand\xG{\mathcal{A}} 
\newcommand\xV{\mathbb{A}} 
\newcommand\xall{\mathfrak{A}}
\newcommand\xA{A} 
 
\newcommand\xL{L} 
\newcommand\xGL{{\xG^L}} 
\newcommand\xVL{{\xV^L}} 
\newcommand\xl{l} 
\newcommand\xg[1][\xl]{{g_{#1}}} 
\newcommand\xa[1][\xl]{{a_{#1}}} 
\newcommand\xb[1][\xl]{{b_{#1}}} 
\newcommand\xx[1][\xl]{{x_{#1}}} 
\newcommand\xy[1][\xl]{{y_{#1}}} 
\newcommand\xv[1][]{{v_{#1}}} 
\newcommand\xw[1][]{{w_{#1}}} 
\newcommand\xgL{\bm{g}}
\newcommand\xhL{\bm{h}}
\newcommand\xaL{{\bm{a}_\xL}}
\newcommand\xbL{{\bm{b}_\xL}}
\newcommand\xxL{\bm{x}}
\newcommand\xyL{\bm{y}}
\newcommand\xuL{\bm{u}}
\newcommand\xvL{\bm{v}}
\newcommand\xwL{\bm{w}}
\newcommand\xD{\Delta}

\newcommand\xF[1]{{F_{#1}}}         
  
\newcommand\xbet{\beta}            
            
\newcommand\xbets{\beta^*}         
\newcommand\xbetc{\beta_\text{c}}         
\newcommand\xbetb{\hat\beta}

\newcommand\xGam[3]{\Gamma_{{#1}{#2}}\p{{#3}}}
\newcommand\xgam[3]{\Gamma'_{{#1}{#2}}\p{{#3}}}
\newcommand\xggam[3]{\Gamma''_{{#1}{#2}}\p{{#3}}}
\newcommand\xGamL[3]{\bm{\Gamma}_{{#1}{#2}}\p{{#3}}}
\newcommand\xgamL[3]{\bm{\Gamma}'_{{#1}{#2}}\p{{#3}}}
\newcommand\xggamL[3]{\bm{\Gamma}''_{{#1}{#2}}\p{{#3}}}
\newcommand\xgams{{\bm{\Gamma}^{\prime*}}}
\newcommand\xggams{{\bm{\Gamma}^{\prime\prime*}}}
\newcommand\xr{r}
\newcommand\xs{s}
\newcommand\xt{t}
\newcommand\xnr{{\bar r}} 
\newcommand\xns{{\bar s}}

\newcommand\xwalks[2]{\bm{Z}_{{#1}{#2}}}
\newcommand\xwalksq[2]{\tilde{\bm{Z}}_{{#1}{#2}}}
\newcommand\xwalksqrestrict[2]{\tilde{\bm{Z}}_{{#1}{#2}}^{\prime}}

\newcommand\xdist[2]{{d_{{#1}{#2}}}}
\newcommand\xdistL[2]{{\bm{d}_{{#1}{#2}}}}

\newcommand\xmeanl[1]{{\left\langle{#1}\right\rangle}_{\xl}}
\newcommand\xmeanls[1]{{\left\langle{#1}\right\rangle}_{\xa,\xb}^{\xp{}{}}}
\newcommand\xmeanp[2]{{\left\langle{#2}\right\rangle}^{s,r}_{\xx,\xy}}

\newcommand\xp[2]{p_{{#1}{#2}}}

\newcommand\xAp[3]{{\left(\xA^{#3}\right)_{{#1}{#2}}}}

\newcommand\xAe[3]{{\left(\e^{{#3}\xA}\right)_{{#1}{#2}}}}
\newcommand\xAed[3]{{\left(\xA\e^{{#3}\xA}\right)_{{#1}{#2}}}}

\newcommand\xN{N}

\newcommand\xn[1][]{{n_{#1}}}

\newcommand\xmartFs[3]{{\mathfrak{M}^*({#1},{#2},{#3})}}
\newcommand\xmartF[3]{{\mathfrak{M}({#1},{#2},{#3})}}

\title{Accessibility Percolation on Cartesian Power Graphs}

\author{Benjamin Schmiegelt\thanks{\href{mailto:schmiegb@thp.uni-koeln.de}{schmiegb@thp.uni-koeln.de}} \and Joachim Krug\thanks{\href{mailto:jkrug@uni-koeln.de}{jkrug@uni-koeln.de}} }
\institute{
    B. Schmiegelt \email{schmiegb@thp.uni-koeln.de} \and J. Krug \email{jkrug@uni-koeln.de}
        \at Institute for Biological Physics, University of Cologne
}

\journalname{}
\date{}

\begin{document}

\maketitle

\begin{abstract}
A fitness landscape is a mapping from a space of discrete genotypes to the
real numbers. A path in a fitness landscape is a sequence of genotypes
connected by single mutational steps. Such a path is said to be
accessible if the fitness values of the genotypes encountered along
the path increase monotonically. We study accessible paths on random fitness
landscapes of the House-of-Cards type, on which fitness
values are independent, identically and continuously distributed
random variables. The genotype space is taken to be a Cartesian power
graph $\xGL$, where $\xL$ is the number of genetic loci and the
allele graph $\xG$ encodes the possible allelic states and
mutational transitions on one locus. The probability of existence of
accessible paths between two genotypes at a distance linear in $\xL$
displays a transition from 0 to a positive value at a threshold $\xbetc$
for the fitness difference between the initial and final genotype. We
derive a lower bound on $\xbetc$ for general $\xG$ and
show that this bound is tight for a large class of allele
graphs. Our results generalize previous results for accessibility
percolation on the biallelic hypercube, and compare favorably to published
numerical results for multiallelic Hamming graphs.
\end{abstract}

\section{Introduction}

\begin{figure}
    \begin{center}
    \hfill
        \fbox{
		\includegraphics{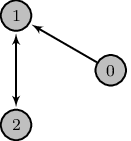}
	}
	\hfill
        \fbox{
		\includegraphics{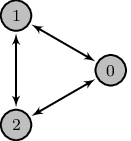}
	}
	\hfill .
    \end{center}
    \begin{center}
        \fbox{
		\includegraphics{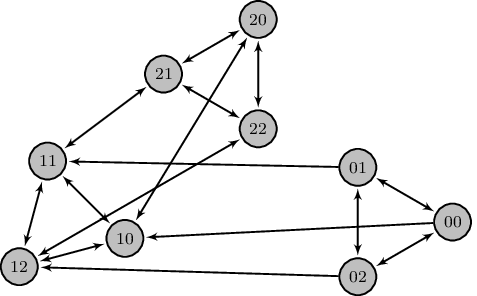}
	}
    \end{center}
    \caption{
        Example of a genotype space as the Cartesian graph product of two allele graphs.
        The two allele graphs are shown on top with the genotype space below.
        While the second factor graph represents a locus with three possible alleles, all of which may mutate freely from one to another, the first factor graph represents a locus on which not all mutations between the alleles are considered possible.
        Specifically mutations between $0$ and $2$ must take an intermediate mutation through $1$ and additionally the mutation from $0$ to $1$ is considered irreversible and does not allow backstepping to $0$.
        Although in this work we define the genotype graph as the direct Cartesian power of a single allele graph, different allele graphs as shown here can still be modeled without loss of generality by assuming that $\xG$ is the disjoint sum of the individual graphs.
        Since the individual constituent graphs are not connected in this sum, this does not increase
        the accessibility (see Remark \ref{remark:alleles}) \label{Fig:Cartesian}
    }
\end{figure}

\begin{figure}
    \begin{center}
        \fbox{
		\includegraphics{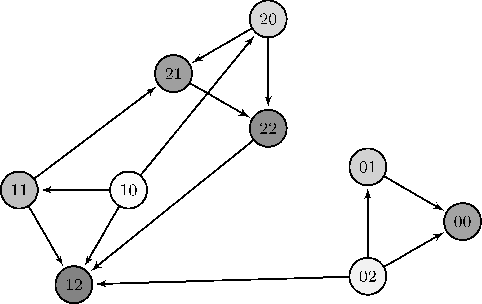}
	}
    \end{center}
    \caption{
        Example of a fitness graph generated from the genotype space in Figure \ref{Fig:Cartesian} according to the HoC model.
        The opacity of nodes indicates the randomly chosen fitness value.
        Only arrows representing mutations that were originally allowed in the genotype graph and also point towards increasing fitness remain, resulting in an acyclic directed graph on genotypes.
        The global minimum and maximum in this realization are $(10)$ and $(12)$ respectively and the latter is accessible from the former by multiple accessible paths, for example the direct one $(10) \rightarrow (12)$, but also $(10) \rightarrow (20) \rightarrow (22) \rightarrow (12)$.
        As a counter-example to accessibility consider $(01)$ and $(12)$.
        Although $(12)$ has higher fitness than $(01)$, it is not accessible from $(01)$ \label{Fig:Fitnessgraph}
    }
\end{figure}

In the strong-selection weak-mutation (SSWM) regime evolutionary dynamics reduces to an adaptive walk on what is known as a fitness landscape, the map from genotypes to fitness values \cite{Fragata2019,deVisser2014}.
For low mutation rates the nearly monomorphic population can be represented by a single majority genotype moving through the space of genotypes by individual mutations that fix with a probability depending on the fitness of the mutant relative to the parental genotype \cite{Gillespie1984,Orr2002}.
Under strong selection, the movement of such a walker is additionally constrained towards increasing fitness values, making it an adaptive walk \cite{Kauffman1987}.
This limits the number of selectively \textit{accessible paths} a population can take through the genotype space \cite{Carneiro2010,Franke2011,Weinreich2005}.

Here we investigate the impact that the mutational structure of the genotype space has on the number of evolutionary paths available to SSWM dynamics.
We use a simple stochastic model for fitness landscapes known as the
House-of-Cards (HoC) model, in which each genotype $\xgL$ is assigned an i.i.d. continuous random fitness value $\xF{\xgL}$ \cite{Kauffman1987,Kingman1978}.
Continuity of the distribution assures that ties in fitness values almost surely do not happen.
Then a path is accessible if these fitness values are in increasing order.
Accessibility therefore is a property purely of the ordering of the i.i.d. random variables.
As a consequence accessibility is independent of the actual distribution chosen and we are free to choose any representative distribution.
We will use the standard uniform distribution which has certain properties that make it easier to work with.

A genotype is made up of many individual sites or loci, which can be found in some given number of states called \textit{alleles}
and can be mutated individually.
For simplicity we will assume that all loci have the same set of possible states.
Therefore genotypes are sequences $\xgL = \p{ \xg[1], \ldots, \xg[\xL] }$, with $\xL$ determining the number of loci.
Individual (point) mutations, which are the only ones to be considered here, mutate only one of the loci. The mutational structure of the system
determines whether every state of one locus is able to mutate to any other or whether some restrictions apply.
For example, whereas point mutations in the DNA sequence can mutate
any nucleotide base into any other, the genetic code constrains the
possible one-step transitions
between amino acids. To accommodate general mutational structures we describe the loci by a simple directed graph.
\begin{definition}
	An \textit{allele graph} is a finite simple directed graph.
\end{definition}
We will denote the allele graph under consideration $\xG$, its vertex set $\xV$, the size of its vertex set $\xall$ and its adjacency matrix $\xA$.
The vertex set $\xV$ is the set of all alleles and the graph's arrows indicate possible one-step mutations between alleles.
The assumption of finiteness is not strictly necessary but allows for more focused proofs.
Extension to infinite graphs is straight-forward with sufficient regularity properties (e.g. bounded degrees).
Again for simplicity, we assume the allele graph to be the same on all loci with the vertices identified by natural numbers.
We will give a justification for this restriction after introduction of the genotype space in the following paragraph.

\begin{definition}
	A \emph{genotype space (over $L$ loci with allele graph $\xG$)} is the Cartesian power graph $\xGL = \xG^{\square\xL}$ \cite{Martinsson2018}, where the Cartesian product $\mathcal{U}\square\mathcal{V}$ of two directed graphs $\mathcal{U}$ and $\mathcal{V}$ is a graph over the Cartesian product of their vertex sets with an arrow from $(u,v)$ to $(u',v')$ iff either there is an arrow from $u$ to $u'$ in $\mathcal{U}$ or there is an arrow from $v$ to $v'$ in $\mathcal{V}$, but not both.
\end{definition}
	With this definition the genotype space's vertex set forms the genotypes, with arrows between genotypes that can be reached via one-step mutations (Figure \ref{Fig:Cartesian}).
The fitness landscape constrains which of these arrows may be taken by an adaptive walker and we call the directed sub-graph of the genotype graph obtained by removing arrows which do not point towards increasing fitness the fitness graph \cite{Crona2013}
(Figure \ref{Fig:Fitnessgraph}).
For conciseness we usually do not specify the number of loci and the allele graph for the genotype space under consideration, assuming these parameters to be $\xL$ and $\xG$ instead.

\begin{remark}
  \label{remark:alleles}
Although we assume the allele graph to be the same on all loci, this still implicitly covers situations in which different loci use different numbers of alleles, as long as the number of alleles is bounded by a constant.
This can be seen by considering an allele graph constructed as the disjoint union of all allele graphs with size at most the constant bound.
The resulting genotype space then also separates into many disjoint components, one of which is the original genotype space with the varying sequence of allele graphs.
Because mutations between the components are impossible, we can consider, without loss of generality, this larger genotype space without affecting the accessibility property.
\end{remark}

\begin{definition}
	A \emph{fitness landscape} on a genotype space $\xGL$ is an assignment of real numbers $\xF{\xgL}$, called \emph{fitness values}, to each genotype $\xgL$.
	The \emph{fitness graph} of the fitness landscape is the directed graph over all genotypes with an arrow from $\xgL$ to $\xhL$ iff there is an arrow from $\xgL$ to $\xhL$ in $\xGL$ and $\xF{\xhL} > \xF{\xgL}$.
\end{definition}

As seen in this definition we will use bold face for genotypes, i.e.\ vertices of the genotype space, while using normal face for alleles, i.e.\ vertices on the allele graph.
Throughout, quantities defined over the allele graph will be written in normal face and analogous quantities defined over the genotype space will be written in bold face.

\begin{definition}
	A \emph{House-of-Cards (HoC) model} over a genotype space $\xGL$ is a random distribution over the set of all fitness landscapes on $\xGL$, such that each fitness value is chosen i.i.d.\ from a standard uniform distribution.
\end{definition}

\begin{definition}
	Given genotypes $\xgL$ and $\xhL$ on a fitness landscape over a genotype space $\xGL$,
	\begin{enumerate}
		\item  a walk on $\xGL$ from $\xgL$ to $\xhL$ is called \emph{accessible} if it is also a walk on the fitness graph of the fitness landscape and
		\item $\xhL$ is said to be \emph{accessible} from $\xgL$ if at least one such walk exists.
	\end{enumerate}
\end{definition}

We write $\xwalks{\xgL}{\xhL}$ for the number of accessible walks from $\xgL$ to $\xhL$, considering it as a random variable over the HoC distribution.

\begin{definition}
	The \emph{(HoC) accessibility} of a genotype $\xhL$ from a genotype $\xgL$ on a genotype space is the probability that $\xhL$ is accessible from $\xgL$ in a HoC model over the genotype space.
\end{definition}
In other words accessibility of $\xhL$ from $\xgL$ is the quantity $\Prob{\xwalks{\xgL}{\xhL} \geq 1}$.

Our goal is to determine the accessibility between pairs of genotypes $\xaL$ and $\xbL$ defined on genotype spaces with $\xL$ loci as $\xL$ becomes large (with fixed allele graph).
This question is in particular non-trivial if the directed distance $\xdistL{\xaL}{\xbL}$ from $\xaL$ to $\xbL$ on the genotype space is of linear order in $\xL$.
Here we understand \emph{directed distance} $\xdistL{\xgL}{\xhL}$ between two genotypes $\xgL$ and $\xhL$ as the length of the shortest (not necessarily accessible) path from $\xgL$ to $\xhL$ on the genotype space.
Of special importance for this question is the value $\xbet$, which we define as the fitness difference $\xF{\xbL}-\xF{\xaL}$.
\begin{definition}
	For values $0 \leq \xbet \leq 1$, the \emph{$\xbet$-(HoC-)accessibility} of a genotype $\xhL$ from a genotype $\xgL$ on a genotype space is the probability that $\xhL$ is accessible from $\xgL$ in the HoC model, given that the HoC distribution is conditioned on $\xF{\xhL}-\xF{\xgL} = \xbet$.
\end{definition}
We typically still write $\Prob{\xwalks{\xhL}{\xgL} \geq 1}$ to refer to $\xbet$-accessibility.
If there is ambiguity between the two in a given context, we add notation to indicate the conditioning.

It is known from previous work on the case of two alleles, $\xV =
\{0,1\}$, and linear distance $\xdistL{\xaL}{\xbL} \sim \delta \xL$,
that there is a critical value $\xbetc$, depending on $\delta$, such that for constant choices of $\xbet$ above or below $\xbetc$, asymptotically the probability of $\xbL$ being $\xbet$-accessible from $\xaL$
converges to $1$ or $0$, respectively, as $\xL\rightarrow\infty$
\cite{Berestycki2016,Berestycki2014,Hegarty2014,Li2018,Martinsson2015}.
The transition occurring at $\xbet = \xbetc$ has been referred to as \textit{accessibility percolation} \cite{Krug2019,Nowak2013}.
Apart from a computational study \cite{Zagorski2016}, so far accessibility percolation has been studied only for the biallelic case for which the genotype space $\xGL$ is the $\xL$-dimensional (binary) hypercube.

\begin{remark}
Results related to those
presented here have been obtained in the context of first-passage
percolation \cite{Kistler2020,Martinsson2016,Martinsson2018}. The accessibility percolation problem (with any continuous distribution) can be mapped to an equivalent first-passage percolation problem with uniformly distributed weights as described in \cite{Martinsson2015}.
This mapping does however map accessibility percolation with fitness values on vertices to first-passage percolation with weights on vertices as well, while traditionally weights are put on edges in that context.
In \cite{Martinsson2018} Martinsson considers a first-passage percolation model that would map to the HoC accessibility percolation problem if fitness values were assigned to edges rather than vertices.
We will adapt and extend his methods to directly resolve the specific accessibility problem introduced here without requiring the mapping to first-passage percolation.
\end{remark}

\begin{remark}
Another way of looking at the HoC $\xbet$-accessibility problem is to consider it as a Bernoulli percolation on a certain ensemble of orientations of $\xGL$.
Alternatively to the definition of $\xbet$-accessibility above, the $\xbet$-conditioning in the HoC model can also be applied by conditioning on $\xF{\xaL} = 0$ and $\xF{\xbL} = \xbet$.
This follows immediately from the fitness values being i.i.d.\ uniform random variables.
If all fitness values are increased by a constant value, taking their remainder modulo $1$, the resulting distribution is unchanged, so that the $\xbet$-accessibility must be unaffected by further conditioning on the initial fitness value.
With this then only genotypes with fitness values below $\xbet$ are relevant in determining whether an accessible path exists.
Furthermore, after removal of the ineligible genotypes, accessibility will depend only on the order of the fitness values on the remaining vertices.
Because the fitness values are chosen i.i.d. this implies that $\xbet$-accessibility in the original problem is equivalent to $1$-accessibility after additional removal of each vertex (that is not $\xaL$ or $\xbL$) with probability $1-\xbet$, i.e. a Bernoulli site percolation with rate $\xbet$.
This perspective may be more suitable if one is to consider the effect of decreasing $\xbet$.
\end{remark}

\section{Results}

In this section we present our general results for (mostly) arbitrary allele graphs $\xG$ of which specific applications will be demonstrated in Section \ref{Sec:Applications}.
First we define some attributes of the problem more carefully and state the limitations imposed by our subsequent proofs.

\subsection{Prerequisites}

Our intent is to describe the limiting behavior of accessibility as $L\rightarrow\infty$ for fixed allele graphs between genotypes in distances which are large, i.e.\ linear in $\xL$.
To make this setup precise we need to introduce some more notation.
We also introduce some restrictions on the choices of the allele graph $\xG$ and the choices for the sequence of endpoints $\xaL$ and $\xbL$ in order to avoid pathological and non-converging behavior.
These restrictions and assumptions are summarized in Definition \ref{Def:Setup}.

First we note that since allele graphs were defined to be finite, there is a maximal degree $\xD < \infty$ among all vertices.
When extending the results to infinite allele graphs it should be required that such an upper bound on the degree still exists in order for most results to remain valid.
If the degrees are not sufficiently bounded in an infinite allele graph, then the number of walks from $\xaL$ to $\xbL$ may become so large that the problem results in trivial accessibility.

As $\xL$, and with it the graph $\xGL$, changes we need to define a sequence of endpoint pairs $(\xaL,\xbL)$ in $\xL$.
Correspondingly we also intend $\xbet$ to vary with $\xL$ as we consider $\xbet$-accessibility.
For convenience the dependence of $\xbet$ on $\xL$ is taken to be implicit and usually not reflected in notation.
For general sequences of endpoints calculations may become tediously complex and so we impose a few restrictions described in the following on the sequences for which our results will apply.
Because the position of a locus in the sequence of the genotypes does not matter, all relevant properties of the pair $(\xaL,\xbL)$ for a given $\xL$ can be expressed as an integer-valued matrix:
\begin{definition}
	The \emph{allele counting matrix} of a pair of genotypes $(\xaL,\xbL)$ over a genotype space is the $(\xall,\xall)$-matrix with entries
\begin{align}
    M_{\xv\xw} = \left| \{\xa = \xv \land \xb = \xw \;|\; \xl=1\ldots\xL\}\right|
\end{align}
where $\xv,\xw\in\xV$ are alleles.
\end{definition}
This matrix counts for each pair of alleles the number of loci on which the path is required to move from $\xv$ to $\xw$, thereby dividing out the permutation-symmetry of loci.
A sequence of such matrices $M$ in $\xL$ is equivalent to a sequence of pairs $(\xaL,\xbL)$ up to the irrelevant symmetry of $\xGL$.

For each value of $\xL$ we have then a matrix $M$ corresponding to the pair $(\xaL,\xbL)$ of endpoints.
Because we are interested in the behavior as $\xL$ diverges, we want to focus on cases where this sequence of matrices is sufficiently well-behaved. Borrowing the terminology from phylogeny studies \cite{Lockhart1994}, 
we introduce the following object:
\begin{definition}
	The \emph{divergence matrix} for a sequence of pairs of genotypes $((\xaL,\xbL))_{\xL}$ with allele counting matrices $M^{(\xL)}$ on a corresponding sequence of genotype spaces with $\xL$ alleles and fixed allele graph is the $(\xall,\xall)$-matrix with elements
\begin{align}
  \label{Eq:pmatrix}
	\xp{\xv}{\xw} = \lim_{\xL\rightarrow\infty}\frac{M^{(\xL)}_{\xv\xw}}{\xL}
\end{align}
for all $\xv,\xw\in\xV$, assuming all limits exist.
	We also define a sequence of \emph{divergence remainder matrices} with elements
	\begin{align}
		R^{(\xL)}_{\xv\xw} = M^{(\xL)}_{\xv\xw} - \xL\xp{\xv}{\xw}.
	\end{align}
\end{definition}
We want this divergence matrix to exist and the remainder terms to be sufficiently well-behaved that we can make convergence statements about the accessibility question.
Specifically we require the following:
\begin{definition}
  \label{Def:Setup}
	A \emph{(well-behaved) accessibility setup} is a sequence $(\xGL)_\xL$ of genotype spaces with $\xL$ loci on the same allele graph $\xG$ together with a sequence of pairs of genotypes $((\xaL,\xbL))_\xL$ on these genotype spaces such that
	\begin{enumerate}
		\item for all $\xL$: $M^{(\xL)}_{vw} = 0$ whenever there is no non-zero length walk from $v$ to $w$ in $\xG$,
		\item for all $\xL$: $\xaL \neq \xbL$,
		\item the divergence matrix $p$ of $((\xaL,\xbL))_\xL$ exists,
		\item at least one off-diagonal element of $p$ is non-zero.
	\end{enumerate}
\end{definition}
The first requirement assures that there is always at least one candidate path from $\xaL$ to $\xbL$, so that the trivial case does not need to be considered and that it is actually possible to move on all loci.
If there are only zero-length walks on a locus (i.e. start and end point are the same without any possibility of mutation), then effectively the locus cannot affect accessibility in any way, resulting in a pathological effective reduction of $\xL$.

The second requirement assures that we do not need to consider the pathological case in which the $\xbet$-conditioning cannot be satisfied.

The third requirement assures that the ``direction'' between the pair of endpoints is asymptotically well-behaved without oscillations which would need to be reflected in the critical $\xbet$.

The last requirement assures that the distance between the endpoints grows linearly in $\xL$, which is the only limit we consider here.

In the following we are always working with one such implied accessibility setup.
In order to succinctly state our results, we introduce the following quantity for pairs of alleles $\xv,\xw\in\xV$ and $\xt > 0$:
\begin{align}
    \xGam{\xv}{\xw}{\xt} &= \ln\p {\xAe{\xv}{\xw}{\xt}},
\end{align}
with $\xA$ the adjacency matrix of the allele graph.
The exponential is a matrix exponential from which the element representing alleles $(\xv,\xw)$ is extracted, rather than the exponential of the element of the matrix.
In addition to the quantity $\xGam{\xv}{\xw}{\xt}$, also its first two derivatives $\xgam{\xv}{\xw}{\xt}$ and $\xggam{\xv}{\xw}{\xt}$ with respect to $\xt$ will be important.
We indicate the derivative with respect to the $\xt$ argument by backticks as shown.

\begin{proposition}
	\label{CorGam}
	$\xGam{\xv}{\xw}{\xt}$ on $\xt \geq 0$ has the following properties:
	\begin{enumerate}
		\item If $\xv \neq \xw$ and there exists no walk from $\xv$ to $\xw$ on the allele graph, it is nowhere defined (or $-\infty$ everywhere),
		\item if $\xv = \xw$ and there exists no walk of non-zero length from $\xv$ to $\xw$ on the allele graph, it is $0$ everywhere,
		\item otherwise it is strictly monotonic increasing, converging to positive infinity as $\xt\rightarrow\infty$ and to negative infinity as $\xt\rightarrow 0$.
	\end{enumerate}
In case 1. we formally interpret $\e^{\xGam{\xv}{\xw}{\xt}}$ as $\xAe{\xv}{\xw}{\xt}$, which would be zero everywhere.

\end{proposition}
\begin{proof}
	These properties are direct consequences of the behavior of the matrix exponential. \hfill \qed
\end{proof}

We then define the same quantity for pairs of genotypes $\xvL,\xwL\in\xVL$ as averages over the per-locus quantity:
\begin{align}
    \label{DefGamL}
    \xGamL{\xvL}{\xwL}{\xt} &= \xmeanl{\xGam{\xv[\xl]}{\xw[\xl]}{\xt}}
\end{align}
where we use the general notation
\begin{align}
  \xmeanl{X_l} = \frac{1}{\xL}\sum_{\xl=1}^\xL X_l
 \end{align}
 to mean an average over $l$ of the term $X_l$ containing $l$ as a variable.
As we already did for genotypes, quantities acting on $\xGL$ will be written in boldface, while equivalent quantities acting on a single $\xG$ copy will be denoted in normal font-face.
The canonical connection between the former and latter is averaging over loci.
In particular, $\xGamL{\xvL}{\xwL}{\xt}$ can be interpreted as the exponential rate with which the expected number of accessible paths grows in $\xL$, or more precisely the number of quasi-accessible walks, a concept which will be introduced in section \ref{Sec:UpperBound}.

\begin{proposition}
	\label{CorGamL}
	In a (well-behaved) accessibility setup $\xGamL{\xvL}{\xwL}{\xt}$ satisfies the following properties on $\xt > 0$:
	\begin{enumerate}
		\item $\xGamL{\xvL}{\xwL}{\xt}$ is strictly monotonic increasing in $\xt$, diverging to $\infty$ as $\xt\rightarrow \infty$ and to $-\infty$ as $\xt\rightarrow 0$,
		\item $\xGamL{\xvL}{\xwL}{\xt}$ is bounded from above by a continuous strictly monotonic increasing function of $\xt$,
		\item $\xgamL{\xvL}{\xwL}{\xt}$ is bounded from below and above by continuous strictly positive functions of $\xt$.
		\item All higher derivatives of $\xGamL{\xvL}{\xwL}{\xt}$ are bounded from below and above by continuous functions.
	\end{enumerate}
\end{proposition}
\begin{proof}
	The conditions on the allele counting matrix in an accessibility setup guarantee that all contributions to the mean in eq. (\ref{DefGamL}) fall under the last point of Proposition \ref{CorGam}.
	Furthermore because the allele graph is finite and fixed, there is only a finite number of pairs $(\xv, \xw)$ of loci.
	The properties in this proposition then follow immediately from $\xGamL{\xvL}{\xwL}{\xt}$ being a point-wise average of a finite number of continuous strictly monotic increasing funtions with the same properties and only the relative weighting dependent on $\xL$. \qed
\end{proof}
As a consequence of the first property there exists exactly one value for each $\xL$ at which $\xGamL{\xv}{\xw}{\xt}$ becomes $0$, which we denote $\xbetb$.
This is going to be our candidate for the threshold of $\xbet$-accessibility.

We also require the following function with domain $0 \leq \xr,\xs \leq 1$, which is a slight generalization of a function introduced by Martinsson in \cite{Martinsson2018}:
\begin{definition}
	\emph{Martinsson's function} of an accessibility setup is the function on $[0,1]^3$ defined by
\begin{align}
  \xmartF{\xs}{\xr}{\xbet} &= \xmeanl{\xmeanp{}{ \xGam{\xx}{\xy}{\xbet\xs} }},
\end{align}
where $\xnr = 1-\xr$ and $\xns = 1-\xs$ and
$\xmeanp{}{\cdot}$ is the mean over $\xx,\xy\in\xV$ weighted by
\begin{align}
    \e^{\xGam{\xa}{\xx}{\xbet\xns\xr} + \xGam{\xx}{\xy}{\xbet\xs} + \xGam{\xy}{\xb}{\xbet\xns\xnr}}.
\end{align}
\end{definition}
Explicitly written out it reads
\begin{align}
   \label{Eq:Martinssonfunction}
    \xmartF{\xs}{\xr}{\xbet} &=\xmeanl{
    \frac{
        \sum_{\xx,\xy\in\xV}\xGam{\xx}{\xy}{\xbet\xs}\e^{\xGam{\xa}{\xx}{\xbet\xns\xr} + \xGam{\xx}{\xy}{\xbet\xs} + \xGam{\xy}{\xb}{\xbet\xns\xnr}}
    }{
        \sum_{\xx,\xy\in\xV}\e^{\xGam{\xa}{\xx}{\xbet\xns\xr} + \xGam{\xx}{\xy}{\xbet\xs} + \xGam{\xy}{\xb}{\xbet\xns\xnr}}
    }}.
\end{align}
For the case that $\xy$ is not reachable from $\xx$ or that $\xs = 0$, the formula yields negative infinities for $\xGam{\xx}{\xy}{\xbet\xs}$.
We assume that in this case formally the natural choice
\begin{align}
    \xGam{\xx}{\xy}{\xbet\xs}\e^{\xGam{\xx}{\xy}{\xbet\xs}} = 0
\end{align}
holds.
Martinsson's function can be interpreted as a refined version of $\xGamL{\xvL}{\xwL}{\xbet}$ in which walks are segmented into three parts, each of which traverses fitness spans of $\xbet\xns\xr$, $\xbet\xs$ and $\xbet\xns\xnr$ respectively, weighting the expected number of accessible walks on the middle segment by the number of accessible walks by which it can be reached from $\xaL$ and $\xbL$.
In this way a positive value of Martinsson's function relative to $\xGamL{\xaL}{\xbL}{\xbet}$ shows that the number of accessible walks is clustered around few initial and end segments with many alternative accessible middle segments, which in turn implies that the overall expected value is not a good indicator for the existence of at least one accessible walk.
How this function arises from consideration of a variation of the second moment method is described in Section \ref{Sec:LowerBound}.

\begin{proposition}
    $\xmartF{\xs}{\xr}{\xbet}$ satisfies the following properties:
	\begin{enumerate}
		\item $\xmartF{0}{\xr}{\xbet} = 0$,
		\item $\xmartF{1}{\xr}{\xbet} = \xGamL{\xaL}{\xbL}{\xbet}$,
		\item $\xmartF{1}{\xr}{\xbetb} = 0$.
	\end{enumerate}
\end{proposition}
\begin{proof}
The first two statements can be derived immediately by application of matrix multiplication.
The third statement is an immediate consequence of the second given the definition of $\xbetb$. \qed
\end{proof}

The objects introduced so far are dependent on $\xL$ implicitly through the averaging process over loci.
In order to be able to make statements about the limiting behavior, it is useful to consider the limits of these quantities as $\xL\rightarrow\infty$.
We use the non-$\xL$-dependent mean
\begin{align}
  \xmeanls{X_{\xv\xw}} = \sum_{\xv,\xw\in\xV} \xp{\xv}{\xw} X_{\xv\xw}
\end{align}
  as a replacement for $\xmeanl{X_{\xv\xw}}$.
As the number of alleles is finite, $\xmeanl{X_{\xv\xw}}$ will converge to $\xmeanls{X_{\xv\xw}}$ as $\xL\rightarrow\infty$.

In particular we write $\xbets$ for the limit of $\xbetb$ as $\xL \rightarrow \infty$.
Similarly we write for the limit of Martinsson's function
\begin{align}
    \xmartFs{\xs}{\xr}{\xbet} &= \xmeanls{\xmeanp{}{ \xGam{\xx[\xl]}{\xy[\xl]}{\xbet\xs} }}.
\end{align}

\begin{proposition}
	\label{CorBetS}
	In a (well-behaved) accessibility setup
	$\xbets = \lim_{\xL\rightarrow\infty} \xbetb$ exists, is positive and satisfies $\xmeanls{\xGam{\xa}{\xb}{\xbets}} = 0$.
\end{proposition}
\begin{proof}
	Per the requirements on the divergence matrix for an accessibility setup the weights in the averaging over pairs of loci in $\xGamL{\xaL}{\xbL}{\xbet}$ converge.
	Because the average is only over a finite number of terms, it consequently also converges.
	Continuity of the matrix exponential and the properties from Proposition \ref{CorGamL} then immediately prove this proposition.
        \qed
\end{proof}

With the necessary quantities defined we can classify accessibility setups as follows:
\begin{definition}
If not only at $\xs = 0$ and $\xs = 1$, but everywhere in its domain $\xmartFs{\xs}{\xr}{\xbets} \leq 0$, then we say that the accessibility setup is of \emph{semi-regular} type.
Otherwise we say that it is of \emph{irregular} type.
If $\xmartFs{\xs}{\xr}{\xbets} < 0$ holds strictly everywhere except at $\xs = 0$ and $\xs = 1$ and if additionally the derivative $\partial_\xs \xmartFs{\xs}{\xr}{\xbets}$ is not zero at $\xs = 1$, then we say that the accessibility setup is of \emph{regular} type.
\end{definition}

For our statements and in the following proofs we make use of Landau notation with the usual meanings of $\O{\cdot}$, $o\p{\cdot}$, $\omega\p{\cdot}$ and $\Theta\p{\cdot}$.
In our notation of arithmetic terms and equations these symbols are stand-ins for some function in the respective class.
The limit variable to which these symbols apply should be evident from context, but is usually $\xL\rightarrow\infty$.
Functions in these classes are not required to be non-negative.
In particular e.g. $|\Theta\p{1}|$ is used to enforce positiveness of a term that is of constant (non-zero) asymptotic order in $\xL$.
If not stated otherwise, these symbols are assumed to be uniform in the sense that the functions represented depend only on the limit variable and model parameters, but not on other local variables.

\subsection{Statements}
\label{Sec:Statements}

We state our results in terms of (weak) threshold functions defined as follows.
\begin{definition}
	Given a function $f(\xL)$ we say that a sequence $c_\xL$ in $\xL$ is a \emph{$f(\xL)$-threshold function for $\xbet$-accessibility} if for all $g(\xL) = \omega(f(\xL))$
	\begin{enumerate}
		\item $(c_\xL+|g(\xL)|)$-accessibility has a non-zero limit inferior as $\xL\rightarrow\infty$ and
		\item $(c_\xL-|g(\xL)|)$-accessibility has a zero limit superior as $\xL\rightarrow\infty$.
	\end{enumerate}
\end{definition}
	We refer to the first condition as the \emph{upper side} and the second condition as the \emph{lower side} of the threshold.
In other words, $c_\xL$ determines the asymptotic transition from zero accessibility to non-zero accessibility if we condition the fitness difference between initial and final genotype, with a window of uncertainty of the same order as $f(\xL)$.
In particular if $c_\xL$ is a $f(\xL)$-threshold for some $f(\xL) = o(1)$, then the limit of $c_\xL$ is the critical value $\xbetc$.

\begin{remark}
  \label{remark:weak}
The notion of threshold chosen here is weak in the sense that it doesn't imply a transition from zero to one, but only from zero to some non-zero probability.
We do not think that our results are actually restricted to this weak bound and we expect that arguments analogous to those made in \cite{Martinsson2018} may be used to extend our weak threshold result to a strong threshold with $\liminf_{\xL \rightarrow\infty}\Prob{\xwalks{\xaL}{\xbL} > 0} = 1$, but we did not pursue this improvement here.
\end{remark}

We can now state our main theorems:

\begin{theorem}
	\label{Thm1}
	In an accessibility setup with the common notation used in previous definitions and the sequence
	\begin{align}
  	\label{Eq:threshold}
    		c_\xL = \xbetb - {\xgamL{\xaL}{\xbL}{\xbetb}}^{-1}\frac{\ln\xL}{\xL},
	\end{align}
	\begin{enumerate}
		\item
		$c_\xL$ satisfies the lower side condition for a $\frac{1}{L}$-threshold function and
		\item
			$c_\xL$ also satisfies the upper side condition and therefore is a $\frac{1}{L}$-threshold function if the accessibility setup is of regular type.
	\end{enumerate}
\end{theorem}
The second part of this theorem in particular implies that $\xbetc = \xbets$ for the regular type, meaning that $\xbets$ is indeed the critical value.

\begin{theorem}
	\label{Thm2}
	In an accessibility setup of irregular type $\xbetc > \xbets$ if $\xbetc$ exists.
\end{theorem}

In other words, the irregular type does \emph{not} have $c_\xL$ as given in theorem \ref{Thm1} as a threshold function.

Lastly we consider the length of accessible paths at the critical point:

\begin{theorem}
	\label{Thm3}
	In an accessibility setup at the candidate threshold function from Theorem \ref{Thm1}, i.e. with $\xbet = c_L + \frac{\eta}{\xL}$ for some constant $\eta$, the probability that all $\xbet$-accessible paths have length in the interval $\xgamL{\xaL}{\xbL}{\xbetb} \xbetb \xL \pm g(\xL)\sqrt{\xL}$ converges to $1$ for every function $g(\xL) = \omega(1)$.
\end{theorem}

In other words, if there are accessible paths at the candidate threshold function, then they have up to fluctuations of order $\sqrt{\xL}$ length $\xgamL{\xaL}{\xbL}{\xbetb}\xbetb\xL$ and since Theorem \ref{Thm1} guarantees that the candidate is actually the threshold function for regular setups, this then implies that the critical paths in the regular setup are of that length as well.

The lower side of the threshold functions in Theorem \ref{Thm1} can be derived directly from a consideration of the expected number of (quasi-)accessible walks and an application of Markov's inequality. This approach will be explained in
Sect.~\ref{Sec:UpperBound}, where we also introduce the notion of quasi-accessibility as a tool to
simplify the counting of accessible paths.
In addition, the first-moment approach allows us to prove Theorem \ref{Thm3} by consideration of the expected values separated by walk length (Sect.~\ref{Sec:Proof3}).

To prove the upper side of the threshold function, it is necessary to bound a higher moment of the expected number of (quasi-)accessible walks in relation to the mean.
In particular, using a generalized version of the second moment method, it is sufficient to bound moments of the form
$$
  \E{\frac{\xwalks{\xaL}{\xbL}}{\E{\xwalks{\xaL}{\xbL}}}\ln\frac{\xwalks{\xaL}{\xbL}}{\E{\xwalks{\xaL}{\xbL}}}}
$$
to show asymptotic boundedness of the accessibility away from zero.
The evaluation of this expected value will follow the general ideas used by Martinsson in \cite{Martinsson2018} to bound for every given (quasi-)accessible focal walk the number of other (quasi-)accessible walks, through the deviating \emph{arcs} on the focal walk that generate all such other walks.
In the mean taken over $\xx$ and $\xy$ in Martinsson's function \eqref{Eq:Martinssonfunction},
the focal walk is represented by the walk sequence
\begin{align}
  \label{Eq:walk}
  \xa \rightarrow \xx \rightarrow \xy \rightarrow \xb
\end{align}
and the corresponding three $\Gamma$-terms in the weights, while the deviating arcs are represented by the additional term corresponding to $\xx \rightarrow \xy$ over which the average is performed.
Because Martinsson considers a model that corresponds to putting weights on edges rather than nodes,
our calculations need to be adjusted accordingly (see Sect.~\ref{Sec:LowerBound}).

The lower bound on $\xbetc$ for the irregular case in Theorem \ref{Thm2} is again obtained
following an approach used by Martinsson, by considering walks through pairs of edges $(\xxL,\xxL')$ and $(\xyL,\xyL')$, applying Markov's inequality separately, and union bounding the resulting probability to improve on Markov's inequality from the total expected number of (quasi-)accessible walks
(see Sect.~\ref{Sec:ImprovedBound}). 

\subsection{Asymptotic form}

The theorems as stated in the previous section are dependent on $\xbetb$ and $\xmeanl{\cdot}$ averages, which are $\xL$-dependent quantities.
From the definition of an accessibility setup we do however know that $\xbetb$ converges to $\xbets$ and that averages of the form $\xmeanl{\cdot}$ are asymptotically of the form $\xmeanls{\cdot}$, both of which are $\xL$-independent quantities.
Depending on the specific choice of sequence of pairs $((\xaL,\xbL))_\xL$, the rate of convergence for these quantities may however differ and add additional significant terms in the threshold function, which we detail in this section.

\begin{proposition}
	\label{corollary1}
	In Theorem \ref{Thm1} the sequence $c_L$ can be replaced by some sequence
	\begin{align}
          c_\xL = \xbets - {\xgams}^{-1}\frac{\ln\xL + (1+o(1))\sum_{\xv,\xw\in\xV}R_{\xv\xw}\xGam{\xv}{\xw}{\xbets}}{\xL}
	\end{align}
        for a suitable choice of $o(1)$,	
        and under this change in Theorem \ref{Thm3} there is a suitable choice of $o(1)$ so that the interval becomes
	\begin{align}
          \xgams\xbets\xL + (1+o(1))\sum_{\xv\xw} R_{\xv\xw} \p{\xgam{\xv}{\xw}{\xbets} - \frac{\xggams}{\xgams}\xGam{\xv}{\xw}{\xbets}} \pm g(\xL)\sqrt{\xL}
	\end{align}
        with
        \begin{align}
		\xgams = \xmeanls{\xgam{\xa}{\xb}{\xbets}}, \xggams = \xmeanls{\xggam{\xa}{\xb}{\xbets}}. 
	\end{align}
\end{proposition}

\begin{proof}
The leading order of the distance between the endpoints as $\xL\rightarrow\infty$ is given by the sum of off-diagonal terms of $\xp{\xv}{\xw}$:
\begin{align}
    \xdistL{\xaL}{\xbL} = \delta\xL + o\p{\xL}
\end{align}
where $\delta = \sum_{\xv\neq\xw}\xp{\xv}{\xw}$.
The definition of an accessibility setup enforces that $\delta >0$, so the value of $\xbets$ will be positive, i.e. not zero, and then we can expand $\xbetb$ around $\xbets$ in $\xL$, using that all $\Gamma$-terms are bounded by constants from below and above per Propositions \ref{CorGam} and \ref{CorGamL}:
\begin{align}
    0 = \xGamL{\xaL}{\xbL}{\xbetb}
    =
&\mathrel{}\xmeanls{\xGam{\xa}{\xb}{\xbets}}
	+ \frac{1}{\xL}\sum_{\xv,\xw\in\xV}R_{\xv\xw}\xGam{\xv}{\xw}{\xbets} \\
	& +\xgamL{\xaL}{\xbL}{\xbets}\p{\xbetb-\xbets}
	+ \O{\p{\xbetb-\xbets}^2}.    
\end{align}
The term $\xmeanls{\xGam{\xa}{\xb}{\xbets}}$ is zero by definition of $\xbets$ and $\xgamL{\xaL}{\xbL}{\xbets}$ is bounded from below by a positive constant, so solving for $\xbetb$ results in
\begin{align}
\xbetb &= \xbets - (1+o(1))\frac{1}{\xgamL{\xaL}{\xbL}{\xbets}\xL}\sum_{\xv,\xw\in\xV}R_{\xv\xw}\xGam{\xv}{\xw}{\xbets}
\\&= \xbets - (1+o(1))\frac{\xgams^{-1}}{\xL}\sum_{\xv,\xw\in\xV}R_{\xv\xw}\xGam{\xv}{\xw}{\xbets}.
\end{align}
Inserting this into the candidate threshold function \eqref{Eq:threshold} we obtain the alternative threshold function in the proposition.

Similarly, expanding $\xgamL{\xaL}{\xbL}{c_L + \frac{\eta}{\xL}}$ in Theorem \ref{Thm3} gives, up to terms of order $\frac{\ln\xL}{\xL}$ or smaller
\begin{align}
&\xgams + \xggams\p{c_L - \xbets} + \frac{1}{\xL}\sum_{\xv,\xw\in\xV}R_{\xv\xw}\xgam{\xv}{\xw}{\xbets}
   + \ldots \\
	=&\xgams + \p{1+o(1)}\frac{1}{\xL}\sum_{\xv,\xw\in\xV}R_{\xv\xw}\p{\xgam{\xv}{\xw}{\xbets} + \frac{\xggams}{\xgams}\xGam{\xv}{\xw}{\xbets}} + \ldots 
\end{align}
The remaining terms can contribute at most $\ln\xL$ to the walk length, which would be subsumed by the interval length of order $\sqrt{\xL}$. \hfill \qed
\end{proof} 

In general, the candidate critical value does not depend on the non-linear corrections in the behavior of $(\xaL,\xbL)$, but the leading correction to the critical value changes if the non-linear corrections are of order $\ln\xL$ or higher.

\begin{remark}
If $\sum_{\xv,\xw\in\xV}|R_{\xv\xw}| = \O{1}$, then the sequence $c_L$ in Proposition \ref{corollary1} reduces to
\begin{align}
    c_\xL = \xbets - {\xgams}^{-1}\frac{\ln\xL + \O{1}}{\xL}.
\end{align}
\end{remark}
This condition describes the situation in which $\xaL$ and $\xbL$ are, up to discretization error, separated in a well-defined linear ``direction'' of a fixed allele counting matrix as $\xL$ increases.
In this case the $\O{1}$ contribution is irrelevant since $c_\xL$ represents a $\frac{1}{\xL}$-threshold function, so that additional contributions of order $\frac{1}{\xL}$ lie within the threshold window.
In this linear separation case the threshold function is described fully by the two quantities $\xbets$ and $\xgams$, both of which were derived from the averages over the matrix exponential of the allele graph, weighted by the divergence matrix.

\section{Applications}
\label{Sec:Applications}

\begin{figure}
  \begin{center}
    \hfill\fbox{
		\includegraphics{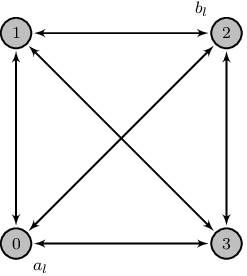}
	}
    \hfill\fbox{
		\includegraphics{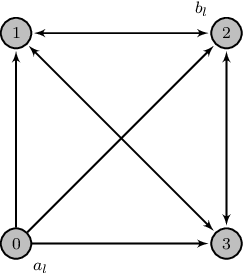}
	}
    \hfill .\\
    \fbox{
		\includegraphics{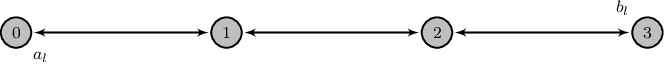}
	}
    \end{center}
	\caption{\label{Fig3}Allele graph structures described in Sect.~\ref{Sec:Applications}. Top left: Complete graph on four alleles with backmutations to the wild-type allele $0$. Top right: Complete graph on four alleles with backmutations to the wild-type allele $0$ removed. Bottom: Path graph on four alleles. In each case the intended initial (wild-type) and final alleles as used in this section are indicated by the labels $\xa$ and $\xb$}
\end{figure}

\subsection{Complete graph}

\begin{figure}
	\includegraphics{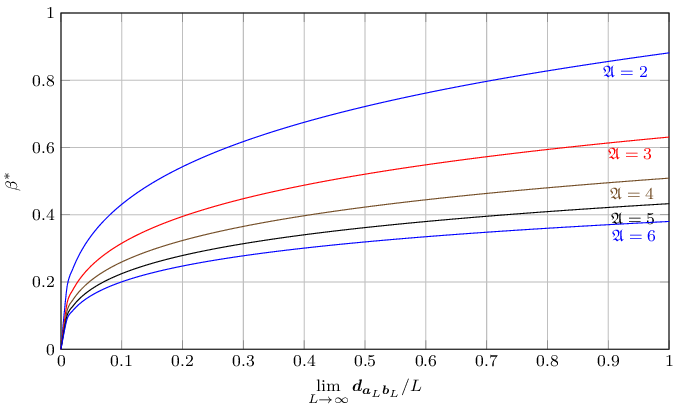}
	\caption{\label{Fig4}
        $\xbets$ as a function of the relative distance $\delta = \lim_{\xL\rightarrow\infty}\frac{\xdistL{\xaL}{\xbL}}{\xL}$ for the complete allele graph with $2-6$ alleles.
        On the complete graph the distance specifies the choice of $\xaL$ and $\xbL$ up to irrelevant symmetries, so that this figure captures the full behavior
    }
\end{figure}

The simplest application is to the complete graph on $\xall = |\xG|$
alleles, as seen in the top-left of Fig.~\ref{Fig3}, which leads to genotype spaces known as Hamming graphs. By
symmetry, in this case there are only two choices for the initial and final allele on a locus, either $\xa = \xb$ or $\xa \neq \xb$.
Therefore the accessibility setup can be fully determined by just the relative distance $\delta$, which is then also the relative Hamming distance.
As shown in \cite{Martinsson2018}, this accessibility setup is (for converging $\delta > 0$) always of regular type for the complete graph.
One obtains
\begin{align}
    \xGamL{\xaL}{\xbL}{\xbet} &= -\ln\xall - \xbet + \delta\ln\p{-1 + \e^{\xall\xbet}} + \bar\delta\ln\p{\xall -1 + \e^{\xall\xbet}},
\end{align}
where $\bar\delta = 1-\delta$. In the biallelic case $\xall = 2$ the
condition $\xGamL{\xaL}{\xbL}{\xbetb} = 0$ reduces to the relation
$\sinh(\xbetb)^\delta \cosh(\xbetb)^{\bar{\delta}} = 1$ which was
first conjectured in \cite{Berestycki2016} and proved in \cite{Li2018,Martinsson2015}.
At full distance $\delta = 1$ without any variation of $\delta$ with $\xL$, $\xbetb=\xbets$ and
\begin{align}
    \xGamL{\xaL}{\xbL}{\xbet} &= -\ln\xall - \xbet + \ln\p{-1 + \e^{\xall\xbet}}.
\end{align}
The values of $\xbets$ and $\xgams$ for small $\xall$ are shown in Table \ref{Table1} and numerically obtained values for $\xbets$ as a function of distance are presented in Fig.~\ref{Fig4}.
\begin{center}
  \begin{table}
    \begin{tabular}{l|l|l|l}
        $\xall$ & $\xbets$ & $\xgams$ & $\xbets\xgams$ \\ \hline
        $2$ & $\arcsin(1) \approx 0.881$ & $\sqrt{2} \approx 1.41$ & $\approx 1.25$ \\
        $3$ & $\ln\p{2\cos\frac{\pi}{9}} \approx 0.631$ & $1+2\cos\p{\frac{2\pi}{9}} \approx 2.53$ & $\approx 1.82$ \\
      $4$ & $\ln\p{\frac{1}{\sqrt{2}}+\sqrt{\sqrt{2}-\frac{1}{2}}} \approx 0.509$ & $\approx 3.60$ & $\approx 1.83$ \\
      $21$ & $\approx 0.154$ & $\approx 20.9$ & $\approx 3.22$
    \end{tabular}
    \caption{Results for the complete allele graph with 2-4 loci, as well as 21 loci, at full distance $\delta = 1$.
      The last column shows the prefactor of the asymptotic walk length at the critical point. In the biallelic case $\xall = 2$
      the result for the walk length was also obtained in \cite{Kistler2020} \label{Table1}}
    \end{table}
\end{center}
In general $\e^{\xbets}$ is the unique positive solution of the
polynomial equation
\begin{align}
    \p{\e^{\xbets}}^{\xall} - \xall \e^{\xbets} - 1 = 0. 
\end{align}
For $\xall \geq 5$ the solution of this equation cannot be expressed in closed form,
however it can be expanded around $\xall \rightarrow \infty$ as
\begin{align}
    \xbets &= \frac{\ln\xall}{\xall} + \frac{1+\ln\xall}{\xall^2} + \O{\frac{\ln\xall}{\xall^3}},
    \\ \xgams &= \xall + \O{\frac{1}{\xall}},
    \\ \xbets\xgams &= \ln\xall + \frac{1+\ln\xall}{\xall} +
                      \O{\frac{\ln\xall}{\xall^2}}. 
\end{align}
As the number of alleles increases, accessibility increases and the
required fitness difference between the start and end point decreases.
In fact this quantity vanishes to zero for $\xall \rightarrow \infty$.
At the same time the length of accessible walks close to the critical
fitness difference increases, but slowly.
The minimal length of a path covering the full distance
$\xdistL{\xaL}{\xbL}$ is $\xL$, and hence $\xbets\xgams -1$ is the fraction
of mutational reversions (where a mutated locus reverts to the
allele it carried in the initial genotype $\xa$) and sideways steps (where a mutation occurs to an allele
that is part of neither the initial nor the target genotype) \cite{Wu2016}.  
The fraction of all alleles on a given locus that appear along an
accessible path close to the critical point is given by $\xall^{-1}
\xbets \xgams$ which decreases with increasing $\xall$.
Zargorski, Burda and Waclaw carried out simulations of this model,
giving $\xbets$ with two digit precision for different values of
$\xall$ \cite{Zagorski2016}.
Their results match the values derived here up to $\pm 0.01$.

\subsection{Complete graph without return to the wild type allele}

We can modify the complete graph slightly to disallow mutations back
to the allele that was present in the initial genotype (the wild type allele), while still allowing mutations between all other allele, see top-right of Fig.~\ref{Fig3}.
In this case the expressions simplify significantly to
\begin{align}
    \xGamL{\xaL}{\xbL}{\xbet} &= \ln\frac{\e^{\p{\xall-2}\xbet}-1}{\xall-2}, \\
  \xbets &= \frac{\ln\p{\xall-1}}{\xall-2},
           \label{directed}
    \\ \xgams &= \xall-1.
\end{align}
The asymptotic behavior for large $\xall$ is the same as for the complete graph.
For $\xall = 2$, the expressions are ill-defined, but the correct expressions coincide with the limits:
\begin{align}
    \xGamL{\xaL}{\xbL}{\xbet} &= \ln\xbet, \\
    \xbets &= 1, \\
    \xgams &= 1. 
\end{align}
In the biallelic case
$\xall = 2$ this describes accessibility percolation on the directed
hypercube, which was considered by Hegarty and Martinsson in
\cite{Hegarty2014}. In this case $\xbets = 1$, which implies that the
directed hypercube is marginally accessible under the HoC model
\cite{Franke2011}.
For the biallelic case not only the critical value, but also the leading order corrections in the threshold function are known \cite{Hegarty2014} and coincide with the order $\frac{\ln\xL}{\xL}$ contribution in our candidate threshold function and the value of $\xgams$ given above.

\subsection{Path graph}

The complete graph is in some sense the best-case scenario for accessibility.
On the opposite side of the spectrum of possible (undirected) allele graphs one can choose the path graph on $\xall$ vertices, shown at the bottom of Fig.~\ref{Fig3}.
In this case the distance between the two end points increases linearly with
the number of alleles and there is a unique order in which mutations on a locus must be applied.
This causes accessibility to become very low.
For $\xall = 2$ the path graph is identical to the complete graph.
However, already for $\xall = 3$ we find
\begin{align}
    \xbets = \frac{\ln\p{3+2\sqrt{2}}}{\sqrt{2}} \approx 1.25.
\end{align}
Since $\xbets$ represents a fitness quantile which must lie between
$0$ and $1$, this value implies that the path graph on three vertices
can never be accessible for any fitness difference if (almost) all
loci need to mutate from one end of the graph to the other.
For higher $\xall$ this effect becomes more pronounced.
As a possible biological application of the path graph the description
of copy-number variants of genes can be mentioned \cite{Altenberg2015}. 

Since the complete graph on two vertices without reversions has
$\xbets = 1$ as shown before in \eqref{directed} and adding edges can
only decrease $\xbets$, it is actually required that the distance
between $\xa[\xl]$ and $\xb[\xl]$ on the allele graph is at least $2$
in order for $\xbets > 1$ to be possible.

\subsection{Example of non-trivial irregular type}

Many graphs seem to be of completely regular type in the sense that no matter which sequence of pairs $(\xaL,\xbL)$ are chosen, the problem is always of regular type.
Martinsson \cite{Martinsson2018} considered different sufficient conditions on graphs to have this property.
But he also lists the smallest graph, of order $4$, which does not have it.
While this example demonstrates that it is possible to have problems of irregular type, it can also be used to generate semi-regular, but not regular, problem types by carefully interpolating the divergence
matrix
\eqref{Eq:pmatrix} between a regular and irregular type pair of alleles.

\begin{figure}
  \begin{center}
	\includegraphics{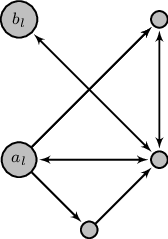}
    \end{center}
    \caption{\label{Fig:IrregularGraph} Example of an allele graph that leads to a problem setup of irregular type with $\xbets < 1$}
\end{figure}

However, the example shown by Martinsson turns out to have $\xbets > 1$, which automatically implies asymptotic inaccessibility 
in the accessibility percolation context due to the defined range of $\xbet = \xF{\xbL}-\xF{\xaL}$ as a
difference of uniform random variables.
We therefore searched for the smallest graph without the regularity property and $\xbets \leq 1$ numerically and found the
example in Fig.~\ref{Fig:IrregularGraph} which has $\xbets \approx 0.983$.

\subsection{Genetic code}

\begin{figure}
    \begin{center}
	\includegraphics{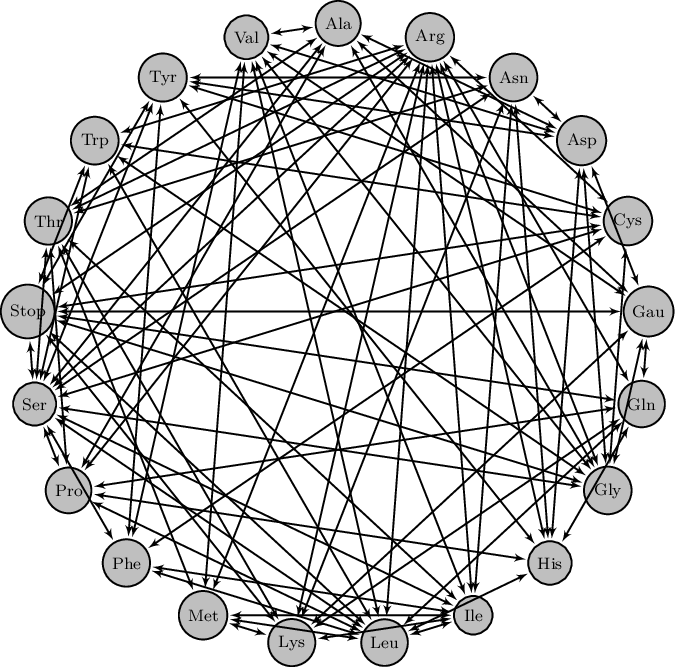}
    \end{center}
    \caption{Allele graph constructed from possible point-mutations on codons. Two amino acids are connected by an arrow iff there is a possible point mutation on a single nucleotide that changes one into the other \label{Fig:codongraph}}
\end{figure}

While the complete graph with $\xall = 4$ may serve as a model for the allele graph of single-nucleotide mutations on DNA or RNA,
the expected effect of such a substitution depends significantly on whether or not it changes the amino acid that is encoded by
the corresponding three-nucleotide codon.
Mutations not affecting the encoded amino acids are known as synonymous. To specifically model the fitness effects
of non-synonymous point mutations we therefore consider the allele graph of all amino acids with edges representing the
mutual reachability by single-nucleotide substitutions (Figure \ref{Fig:codongraph}). 

This graph is considerably less symmetric than the complete graph and in particular the resulting quantities $\xbets$ and $\xgams$ will depend on the particular choices of the path endpoints $\xaL$ and $\xbL$ rather than simply on their distance.
We consider here all pairings of amino acids $\xa[\xl]$ and $\xb[\xl]$, assuming them to be equal for all loci.
Other cases may be interpolated from these.
The results are shown in Table \ref{Table2}.
Whether the given values determine the asymptotic behavior of accessibility exactly depends on whether the regularity criteria
relating to Martinsson's function \eqref{Eq:Martinssonfunction} are satisfied.
Due to the degree of the graph we limited ourselves to numerical tests, which did not indicate any violation of the criteria, although such violations may be more subtle than our tests could verify.

The critical point $\xbets$ and in particular the expected walk length $\xbets\xgams\xL$ are, as one would expect, strongly correlated with the distance between alleles.
The only distance-$3$ pair of amino acids is Tyr/Met which also corresponds to the largest walk length with a
value of $\xbets\xgams \approx 4.7567$.
All other amino acids lie at mutual distance $1$ or $2$.
Nonetheless, the critical point $\xbets \approx 0.4527$ for the distance-$3$ pair lies slightly below that of
the distance-$2$ pair Asp/Met with $\xbets \approx 0.4570$, demonstrating that the overall structure of the allele graph can have a significant impact on accessibility beyond distance.

For comparison, the accessibility of paths
between any pair of codons can be obtained from the values for the complete graph with 4 alleles
(Table \ref{Table1}).
This gives $\xbets \approx 0.51$, while accounting for the multiplication of three bases per codon yields the expected mean critical walk length per codon as $\xbets\xgams \approx 5.5$.
Relative to the complete graph on four alleles the codon graph is allowing arbitrary synonymous mutations without cost, causing the reduction in the critical fitness difference as well as the length of walks at the critical point.
However, when compared to the complete graph on 21 alleles with $\xbets \approx 0.154$ and critical walk length factor $3.22$ (Table \ref{Table1}), which would permit direct mutations between arbitrary amino acids, the fitness cost is still significantly higher on the codon graph, whereas the critical walk length on the codon graph is scattered around that of the 21-allele complete graph.

\begin{table}
    \begin{center}
         \includegraphics[width=\textwidth]{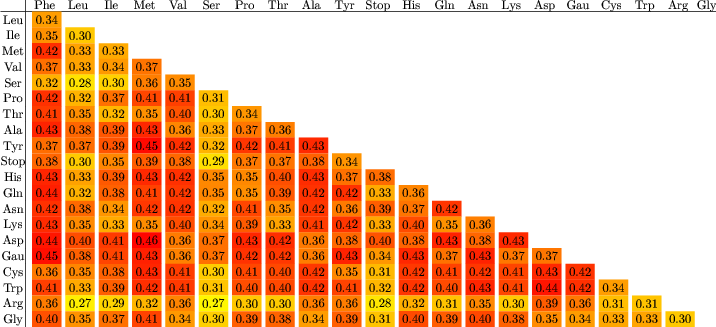}
    \end{center}
    \begin{center}
         \includegraphics[width=\textwidth]{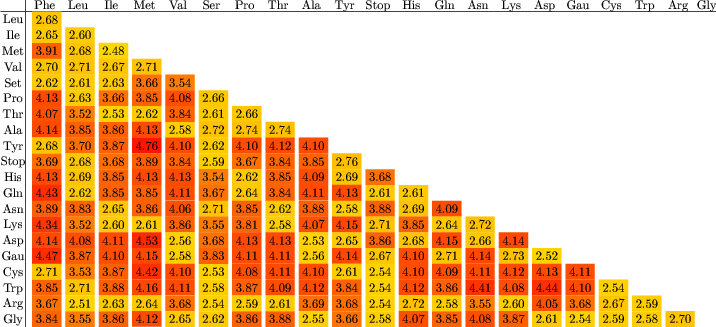}
    \end{center}
    \caption{
      Top: Critical fitness difference $\xbets$ for accessible paths between homopolymer amino acid sequences consisting of the indicated
      pairs. Bottom: Value of $\xgams\xbets$ representing the expected per-locus length of accessible walks at the critical point. All values are obtained numerically and rounded to two digits. Colors represent the magnitude of the displayed values from low (yellow)
      to high (red) (color online) \label{Table2}}
\end{table}

\section{First moment bound and walk length}

\label{Sec:UpperBound}

We start with a proof for the first part of Theorem \ref{Thm1} from an upper bound for accessibility based on the mean number of accessible paths, or rather the mean number of \emph{quasi-accessible} walks.
We define the term \emph{quasi-accessible} as a generalization of the
notion of accessibility used up to now as explained in the following.

\subsection{Quasi-accessibility}

In the original definition of accessibility, a non-self-avoiding walk is never accessible, because
it would have to visit the same fitness value twice, which makes it impossible for the walk to have strictly
increasing fitness.
Handling self-avoidance is non-trivial. To remedy this in a simpler manner,
instead of considering self-avoiding paths on $\xGL$,
we consider an extension of $\xGL$ to $\xGL'$ as follows:
\begin{definition}
	The \emph{extended genotype space} of a genotype space $\xGL$ is the simple directed graph $\xGL'$ with vertex set $\xVL' = \xVL \times \mathbb{N}$ and an arrow from $(\xvL,n)$ to $(\xwL,m)$ iff there is an arrow from $\xvL$ to $\xwL$ in $\xGL$.
\end{definition}
In other words we duplicate every genotype a countable infinite number of times in such a way that traversal of one of its copies can always be replaced by traversal of another copy.
The $1$-section containing all vertices of the form $(\xvL, 1)$ can be identified with the vertices on $\xGL$.
We then assign each of the vertices in $\xGL'$  i.i.d.\ fitness values:
\begin{definition}
	An \emph{extended HoC model} over a genotype space $\xGL$ is the HoC model over the extended genotype space of the genotype space $\xGL$.
\end{definition}
The mentioned $1$-section then corresponds to the original HoC model and we can identify realizations of the extended HoC model with the corresponding realizations of the original HoC model with equal fitness values on the $1$-section.
All other fitness values do not affect this underlying model.
However, it is convenient to introduce these additional fitness values for the following reasons.

We define the following map of walks on $\xGL$ to $\xGL'$.
Each self-avoiding walk is mapped to the corresponding walk on the $1$-section of $\xGL'$.
But instead of mapping non-selfavoiding walks from $\xGL$ to the $1$-section of $\xGL'$, we can
make use of the additional vertex copies to replace all vertices that are visited multiple times in $\xGL$ with
distinct copies in $\xGL'$.
To make this unique, we assume that the $n$-th visit of vertex $\xvL$ in $\xGL$ is mapped to the vertex $(\xvL,n)$ in $\xGL'$,
except if $\xvL$ is the final vertex of the walk, in which case we map the $n$-th visit in reverse order to $(\xv,n)$ in $\xGL'$.
The resulting walk is always selfavoiding in $\xGL'$ and the special case assures that every walk in $\xGL$ is mapped to a walk with endpoints on the $1$-section in $\xGL'$.
\begin{definition}
	A walk on $\xGL$ is \emph{quasi-accessible} if the corresponding mapped walk on $\xGL'$ per the rules above is accessible on $\xGL'$.
	A genotype is said to be quasi-accessible from another if there exists a quasi-accessible walk from the latter to the former.
\end{definition}
\begin{definition}
A walk on $\xGL'$ is $\emph{valid}$ if there is a walk in $\xGL$ which is mapped to it according to the rules above. An (extended) genotype $\xhL'$ on $\xGL'$ is said to be \emph{valid-accessible} from a genotype $\xgL'$ on $\xGL'$ if there exists a valid-accessible walk from the latter to the former on $\xGL'$.
\end{definition}
With these definitions the probability of any walk on $\xGL$, whether self-avoiding or not, to be quasi-accessible is the same, only depending on the length of the walk.
Furthermore the notion of valid-accessibility on $\xGL'$ coincides with quasi-accessibility on $\xGL$.
We denote the number of valid-accessible walks from $(\xvL,1) \in \xVL'$ to $(\xwL,1) \in\xVL'$, or equivalently quasi-accessible walks from $\xvL$ to $\xwL$ with $\xvL,\xwL\in\xVL$, by $\xwalksq{\xvL}{\xwL}$.
Additionally, while quasi-accessibility is different from accessibility for individual walks on $\xGL$, accessibility and quasi-accessibility of one genotype from another on $\xGL$ coincide:
\begin{lemma}
\label{lemma1}
\begin{align}
	\Prob{\xwalks{\xvL}{\xwL} \geq 1} = \Prob{\xwalksq{\xvL}{\xwL} \geq 1}.
\end{align}
\end{lemma}
\begin{proof}
	If $\xwL$ is accessible from $\xvL$, then there exists a walk from $\xvL$ to $\xwL$ which is accessible and therefore also quasi-accessible, implying 
\begin{align}
	\Prob{\xwalks{\xvL}{\xwL} \geq 1} \leq \Prob{\xwalksq{\xvL}{\xwL} \geq 1}.
\end{align}
	If $\xwL$ is quasi-accessible from $\xvL$, then there exists a valid walk on $\xGL'$ from $(\xvL,1)$ to $(\xwL,1)$ which is accessible (with respect to $\xGL'$).
	Removing all vertices $(\xuL,n)$ with $n \neq 1$ from this walk results in another walk on $\xGL'$ which is completely located on the $1$-section due to the validity requirement.
	Because this walk is obtained by only removal of vertices, it is also valid-accessible and because it must be self-avoiding on the $1$-section, it is also an accessible walk from $\xvL$ to $\xwL$ implying the other side of the equality
\begin{align}
	\Prob{\xwalks{\xvL}{\xwL} \geq 1} \geq \Prob{\xwalksq{\xvL}{\xwL} \geq 1}.
\end{align}
\qed
\end{proof}
This implies that we can restrict our investigation to
quasi-accessibility. 

\subsection{Proof of Theorem \ref{Thm1}.1.}

In order to give an upper bound on $\xbet$-quasi-accessibility and with that a proof of Theorem \ref{Thm1}.1, we will consider the mean number of quasi-accessible walks from $\xaL$ to $\xbL$.
Each walk of length $\xN$ from $\xaL$ to $\xbL$ on $\xGL$ is $\xbet$-quasi-accessible with probability
$$\frac{\xbet^{\xN-1}}{(\xN-1)!}$$ where the numerator accounts for the probability that all inner vertices of the walk are found inside the range of fitness values $\xF{\xaL}$ to $\xF{\xbL}$ and the denominator accounts for the increasing order required on these values.
The number of walks taking $\xn$ steps from $\xa[\xl]$ to $\xb[\xl]$ on one locus $\xl$ is given by $\xAp{\xa[\xl]}{\xb[\xl]}{\xn}$.
A walk of length $\xN$ could take each step on any of the loci, so that the total number of walks of length $\xN$ can be written as
\begin{align}
    \sum_{\xn[1]+\ldots+\xn[\xL] = \xN} \binom{\xN}{\xn[1],\ldots,\xn[\xL]} \prod_{\xl=1}^\xL \xAp{\xa[\xl]}{\xb[\xl]}{\xn[\xl]}
\end{align}
where $\binom{\xN}{\xn[1],\ldots,\xn[\xL]}$ is the multinomial
coefficient accounting for the different orderings of steps on individual loci.
Multiplication of this expression with the probability of quasi-accessibility of each such walk gives the mean number of quasi-accessible paths
\begin{align}
    \E{\xwalksq{\xaL}{\xbL}}
    = \sum_{\xN=0}^\infty \sum_{\xn[1]+\ldots+\xn[\xL] = \xN}
  \binom{\xN}{\xn[1],\ldots,\xn[\xL]} \frac{\xbet^{\xN-1}}{\p{\xN-1}!}
  \prod_{\xl=1}^\xL \xAp{\xa[\xl]}{\xb[\xl]}{\xn[\xl]}. 
\end{align}
The term $\p{\xN-1}!$ can be reduced to $\xN!$ by introduction of a derivative
\begin{align}
    \E{\xwalksq{\xaL}{\xbL}}
    = \partial_\xbet \sum_{\xN=0}^\infty \sum_{\xn[1]+\ldots+\xn[\xL] = \xN} \binom{\xN}{\xn[1],\ldots,\xn[\xL]} \frac{\xbet^{\xN}}{\xN!} \prod_{\xl=1}^\xL \xAp{\xa[\xl]}{\xb[\xl]}{\xn[\xl]}
\end{align}
and redistributing all the factorials and $\xbet^\xN$ into the product yields
\begin{align}
    \E{\xwalksq{\xaL}{\xbL}}
    = \partial_\xbet \sum_{\xN=0}^\infty \sum_{\xn[1]+\ldots+\xn[\xL] = \xN} \prod_{\xl=1}^\xL \frac{\xbet^{\xn[\xl]}}{\xn[\xl]!}\xAp{\xa[\xl]}{\xb[\xl]}{\xn[\xl]}.
\end{align}
Finally the sums and the product can be interchanged and
\begin{align}
    \E{\xwalksq{\xaL}{\xbL}}
    &= \partial_\xbet \prod_{\xl=1}^\xL \sum_{\xn=0}^\infty \frac{\xbet^{\xn}}{\xn!}\xAp{\xa[\xl]}{\xb[\xl]}{\xn} 
    = \partial_{\xbet}\prod_{\xl=1}^\xL \xAe{\xa[\xl]}{\xb[\xl]}{\xbet} 
    = \partial_\xbet \e^{\xL \xGamL{\xaL}{\xbL}{\xbet}}
    \\ &= \xL \xgamL{\xaL}{\xbL}{\xbet} \e^{\xL \xGamL{\xaL}{\xbL}{\xbet}}.
\end{align}

We now choose $\xbet$ based on Theorem \ref{Thm1} as
\begin{align}
	\xbet = \xbetb - {\xgamL{\xaL}{\xbL}{\xbetb}}^{-1}\frac{\ln\xL}{\xL} - \frac{|g(L)|}{L}
\end{align}
where $g(L) = \omega(1)$.
Because $\xbetb$ converges to a positive value per Proposition \ref{CorBetS}, around which $\xGam{\xaL}{\xbL}{\xbet}$ and its derivatives are bounded, a Taylor expansion of $\xGamL{\xaL}{\xbL}{\xbet}$ gives
\begin{align}
    \E{\xwalksq{\xaL}{\xbL}}
	= \xL\p{\xgamL{\xaL}{\xbL}{\xbetb} + \mathcal{O}{\p{\xbet-\xbetb}}}
    \e^{\xL \p{\xGamL{\xaL}{\xbL}{\xbetb} + \xgamL{\xaL}{\xbL}{\xbetb}\p{\xbet-\xbetb} + \O{\p{\xbet-\xbetb}^2} } }. 
\end{align}
By definition $\xGamL{\xaL}{\xbL}{\xbetb} = 0$ and by Proposition \ref{CorBetS} $\xgamL{\xaL}{\xbL}{\xbetb}$ converges to a positive value $\xgams$, so that inserting the difference $\xbet-\xbetb$ then gives
\begin{align}
    \E{\xwalksq{\xaL}{\xbL}}
    = (\xgams+o(1))\e^{-\xgams|g(L)|},
\end{align}
implying with $g(L) = \omega(1)$ that
\begin{align}
	\lim_{\xL\rightarrow\infty}\E{\xwalksq{\xaL}{\xbL}} = 0.
\end{align}

Then, by Markov's inequality we have $\Prob{\xwalksq{\xaL}{\xbL} \geq 1} \leq \E{\xwalksq{\xaL}{\xbL}}$, implying that quasi-accessibility converges to zero, which then by Lemma \ref{lemma1} also implies that accessibility of $\xbL$ from $\xaL$ converges to zero, proving part 1 of Theorem \ref{Thm1}. \hfill \qed

\begin{corollary}
	\label{CorA}
	By replacing $-|g(\xL)|$ by a constant value $\eta$ in the above, the expected number of quasi-accessible walks will converge to a non-zero value and increasing $\eta$ allows to arbitrarily increase the limit.
\end{corollary}

\subsection{Proof of Theorem \ref{Thm3}}
\label{Sec:Proof3}

A more refined version of the previous argument can be used to prove Theorem \ref{Thm3}.
Specifically the expected number of $\xbet$-quasi-accessible walks can be separated into intervals of walk lengths.
Let $h_N$ be the expected number of $\xbet$-quasi-accessible walks of length $N$ with
\begin{align}
	\xbet = \xbetb - {\xgamL{\xaL}{\xbL}{\xbetb}}^{-1}\frac{\ln\xL}{\xL} + \frac{\eta}{L}
\end{align}
as in Theorem \ref{Thm3} for some fixed value $\eta$.
This number is an expectation value over realizations of fitness values, but in the following we will consider it as just a number indexed by some number $\xN$ representing the walk length.
Summation of all of these numbers then yields the total expected number of $\xbet$-quasi-accessible walks which we calculated already above:
\begin{align}
    \E{\xwalksq{\xaL}{\xbL}} = \partial_{\xbet} \e^{\xL\xGamL{\xaL}{\xbL}{\xbet}} = \sum_{\xN=1}^\infty h_N.
\end{align}
We can interpret this as the value $\phi(1)$ of the function
\begin{align}
    \phi(z) =
        \partial_{z\xbet}\e^{\xL \xGamL{\xaL}{\xbL}{z\xbet}}
        = \sum_{\xN=1}^\infty h_\xN z^{\xN-1}.
\end{align}
This function can be viewed as an (ordinary) generating function for the sequence $h_N$ shifted by one.
The generating function here is not related to the probability distribution of fitness values, but is rather to be understood as simply a counting tool that separates the total expectation value into slots for different walks lengths using the additivity of the expectation value.

The effect of the derivative $\partial_{z\xbet}$ in the generating function can be reversed by integration of each of the monomials, so that
\begin{align}
    \tilde\phi(z) = 
        \e^{\xL \xGamL{\xaL}{\xbL}{z\xbet}}
        = \sum_{\xN=1}^\infty \frac{\xbet}{\xN} h_\xN z^{\xN}
        = \sum_{\xN=1}^\infty \tilde h_\xN z^\xN
\end{align}
is the generating function of the unshifted $h_\xN$ multiplied by $\frac{\xbet}{\xN}$, which is another sequence that we define as $\tilde h_\xN$.
Normalizing $\tilde\phi(z)$ through division by $\tilde\phi(1) = \e^{\xL \xGamL{\xaL}{\xbL}{\xbet}}$ turns the generating function into a probability generating function over the parameter $\xN$ as random variable and this allows us to apply theorems from probability theory.
Again, this probability is not related to the distribution of fitness values, but is introduced here artificially as a counting tool.
The integrated (probability) generating function factorizes over loci as
\begin{align}
    \frac{\tilde\phi(z)}{\tilde\phi(1)} = \prod_{\xl=1}^\xL
        \frac{
            \e^{\xGam{\xa}{\xb}{z\xbet}}
        }{
            \e^{\xGam{\xa}{\xb}{\xbet}}
        }.
\end{align}
As a consequence the random variable $\xN$ under the generating function's distribution can be written as a sum $\xN = \sum_{\xl=1}^\xL \xn[\xl]$, where $\xn[\xl]$ are independent random variables with probability generating functions
\begin{align}
    \e^{\xGam{\xa}{\xb}{z\xbet} -\xGam{\xa}{\xb}{\xbet}}
    = \frac{
        \sum_{\xn[\xl]=0}^\infty \frac{\xbet^{\xn[\xl]}}{\xn[\xl]!}\xAp{\xa}{\xb}{\xn[\xl]} z^{\xn[\xl]}
    }{
        \sum_{\xn[\xl]=0}^\infty \frac{\xbet^{\xn[\xl]}}{\xn[\xl]!}\xAp{\xa}{\xb}{\xn[\xl]}
    }.
\end{align}
This can be seen going in the reverse direction as the generating function of the sum of independent random variables is the product of the individual generating functions of the summands.
Because the degree of $\xA$ is bounded by $\xD$, $\xAp{\xa}{\xb}{\xn[\xl]} \leq \xD^{\xn[\xl]}$ and the tail of the distribution is dominated by an exponential.
This bound is also independent of the chosen loci $\xa$ and $\xb$ and with the chosen $\xbet$ converging in $\xL$, the central limit theorem applies to the sum $\xN$.
The mean $\xL\mu$ and variance $\xL\sigma^2$ of $\xN$ under this distribution can be obtained from the first and second derivatives of the probability generating function as
\begin{align}
    \mu &= \xbet\xgamL{\xaL}{\xbL}{\xbet} \\
    \sigma^2 &= \xbet\xgamL{\xaL}{\xbL}{\xbet} + {\xbet}^2\xggamL{\xaL}{\xbL}{\xbet},
\end{align}
and the central limit theorem implies that for constants $c>0$:
\begin{align}
    \sum_{|\xN-\mu\xL|\geq c\sigma\sqrt{\xL}} \tilde h_\xN = 2\e^{\xL\xGamL{\xaL}{\xbL}{\xbet}}\Phi(-c)\p{1+o(1)}
\end{align}
where $\Phi$ is the standard normal CDF.
Since the sum's upper bound is asymptotic to $\mu\xL$, for all $\tilde h_\xN$ terms appearing in the sum $\tilde h_\xN \geq \frac{\xbet}{\mu\xL}h_\xN$, so that
\begin{align}
    \sum_{|\xN-\mu\xL|\geq c\sigma\sqrt{\xL}} h_\xN \leq 2\frac{\mu\xL}{\xbet}\e^{\xL\xGamL{\xaL}{\xbL}{\xbet}}\Phi(-c)\p{1+o(1)}.
\end{align}
In particular with the choice of the threshold function of Theorem \ref{Thm3} for $\beta$:
\begin{align}
	\sum_{|\xN-\mu\xL|\geq c\sigma\sqrt{\xL}} h_N \e^{\eta}\leq 2\eta\Phi(-c)\p{1+o(1)}.
\end{align}
This allows one to reduce the mean number of $\xbet$-quasi-accessible walks of length outside the interval $\mu\xL \pm c\sigma\sqrt{\xL}$ to any arbitrarily small value by choosing $c$ large enough.
In other words, if there are $\xbet$-quasi-accessible walks at the suggested threshold function, then they are of length $\mu\xL$ with fluctuations of at most order of $\sqrt{\xL}$.
Since all $\xbet$-accessible walks are also $\xbet$-quasi-accessible walks, the same length constraint then also applies to
accessible paths at the threshold function, completing the proof of Theorem \ref{Thm3}. \hfill $\Box$

\section{Proof of Theorem \ref{Thm2}}

\label{Sec:ImprovedBound}

The upper bound on accessibility obtained from the expected value does not take into account any dependence between walks.
We can improve the bound by including some of the dependencies.
This will make it possible to prove our Theorem \ref{Thm2}, i.e. show that $\xbetc > \xbets$ for irregular types.

Let in this section $0 < \xr < 1$ and $0 < \xs < 1$ be constant.
The intention is to choose them later such that $\xmartFs{\xs}{\xr}{\xbets} > 0$ as application to the irregular type.
This choice is always possible in the irregular case since by continuity $\xmartFs{\xs}{\xr}{\xbets}$ cannot be strictly positive only on the boundaries.

Recall that the parameters $\xr$ and $\xs$ determine the fitness spanned by three segments of each walk with $\xs$ determining the fitness fraction spanned by the middle segment, and $\xr$ determining the distribution of the remaining fitness span onto the first and last segment.
More concretely the intended fitness span of the first segment is $\xbet\xns\xr$, of the second $\xbet\xs$ and the third $\xbet\xns\xnr$, adding up to the full fitness span $\xbet$ that needs to be crossed (see Sect.~\ref{Sec:Statements}, eq.~\eqref{Eq:walk} ). 

For each arrow on the genotype space we can consider the interval formed by the fitness values of the two genotypes incident to it.
If a walk from $\xaL$ to $\xbL$ is accessible, then it contains exactly one arrow with a fitness interval containing
the fitness value $\xbet\xns\xr$.
Let this arrow be $(\xxL,\xxL')$ and let $S^{12}_{\xxL,\xxL'}$ be an indicator variable for this fitness value falling on to the arrow $(\xxL, \xxL')$.
Similarly there is exactly one arrow containing the fitness value $\xbet\p{1-\xns\xnr}$.
Let this arrow be $(\xyL,\xyL')$ and the corresponding indicator $S^{23}_{\xyL,\xyL'}$.
These two arrows segment the walk in the closest possible way according to the intended fitness spans mentioned above.
A walk is accessible only if each of the three segments $\xaL \rightarrow \xxL$, $\xxL' \rightarrow \xyL$ and $\xyL' \rightarrow \xbL$ are accessible.
In the following we refer to these segments as segment $1$, $2$ and $3$ respectively.
To obtain an upper bound on the accessibility of $\xbL$ from $\xaL$ it is therefore sufficient to form a union bound of the form:
\begin{align}
	&\Prob{\xwalks{\xaL}{\xbL} \geq 1}
	\\ \leq&
	\sum_{(\xxL,\xxL'),(\xyL,\xyL')\in\xGL} \Prob{S^{12}_{\xxL,\xxL'} \land S^{23}_{\xyL,\xyL'} \land \xwalks{\xaL}{\xxL} \geq 1 \land \xwalks{\xxL'}{\xyL} \geq 1 \land \xwalks{\xyL'}{\xbL} \geq 1}.
\end{align}

We can now separate the expectation over the fitness of the intermediate genotypes $\xxL$, $\xxL'$, $\xyL$ and $\xyL'$,
\begin{align}
	&\Prob{\xwalks{\xaL}{\xbL} \geq 1}
  \\ \leq&
\sum_{(\xxL,\xxL'),(\xyL,\xyL')\in\xGL} \E{\Prob{\xwalks{\xaL}{\xxL} \geq 1 \land \xwalks{\xxL'}{\xyL} \geq 1 \land \xwalks{\xyL'}{\xbL} \geq 1 | \cdot}},
\end{align}
where the dot in the probability indicates conditioning on the fitness values at the end points. The expectation is over all such fitness values satisfying the conditions $S^{12}_{\xxL,\xxL'}$ and $S^{23}_{\xyL,\xyL'}$.

As a result of this conditioning, the quasi-accessibilities of the three segments mentioned in the equation are negatively dependent, so that the joint probability of the events can be upper bounded by the product of individual probabilities. In order for a segment to be accessible under the conditioning of the fitness values at the end points, all internal fitness values on the segmnent must fall into the fitness range between the end points and the internal fitness values must be increasingly ordered.
Because the fitness values are i.i.d.\ in the HoC model these two properties are independent.
Now suppose we condition on one (or two) of the segments being accessible with any particular choice of accessible walks.
This is equivalent to conditioning all the internal fitness values of these walks in an appropriate manner as well.
The accessibility requires these internal fitness values to be constrained to the fitness range of the segment's end points, which makes them unavailable as internal vertices of the remaining segment(s).
However, all other fitness values are i.i.d.\ and unaffected by this conditioning.
The probability that the remaining segment(s) is (are) then accessible is therefore smaller than if no conditioning of the internal vertices of the other accessible segments had been applied, effectively only removing walks through accessible vertices of the conditioned segments from the set of candidate walks.
For example if $\xwalksq{\xaL}{\xxL}$ is at least $1$, then $\xwalksq{\xxL'}{\xyL} \geq 1$ becomes less likely since the existence of a quasi-accessible walk from $\xaL$ to $\xxL$ implies that some fitness values of other genotypes fall in the range $[\xF{\xaL},\xF{\xxL}]$, excluding them for consideration in the range $[\xF{\xxL'}, \xF{\xyL}]$ required for them to be part of a quasi-accessible walk from $\xxL'$ to $\xyL$.
Consequently:
\begin{align}
    \Prob{\xwalksq{\xaL}{\xbL} \geq 1}
    \leq
    \sum_{\p{\xxL,\xxL'},\p{\xyL,\xyL'}} \E{
        \Prob{\xwalksq{\xaL}{\xxL} \geq 1 | \cdot}
        \Prob{\xwalksq{\xxL'}{\xyL} \geq 1 | \cdot}
        \Prob{\xwalksq{\xyL'}{\xbL} \geq 1 | \cdot}
    }.
\end{align}
Since probabilities lie in $[0,1]$, we can use the upper bound $x \leq x^{1-\alpha}$ for $0< \alpha< 1$ on the middle factor and afterwards we can apply Markov's inequality to all three terms to obtain
\begin{align}
    \label{ref1}
    \Prob{\xwalksq{\xaL}{\xbL} \geq 1}
    \leq
    \sum_{\p{\xxL,\xxL'},\p{\xyL,\xyL'}} \E{
        \E{\xwalksq{\xaL}{\xxL} \geq 1 | \cdot}
        \E{\xwalksq{\xxL'}{\xyL} \geq 1 | \cdot}^{1-\alpha}
        \E{\xwalksq{\xyL'}{\xbL} \geq 1 | \cdot}
    }.
\end{align}
The remaining inner expectation values depend only on the differences of the fitness values that they are conditioned on, not the actual placement of that difference.
We introduce the following quantities:
\begin{align}
    \epsilon_{1} &= \xbet\xns\xr - \p{\xF{\xxL}-\xF{\xaL}}, \\
    \epsilon_{2} &= \xbet\xs - \p{\xF{\xyL}-\xF{\xxL'}}, \\
    \epsilon_{3} &= \xbet\xns\xnr - \p{\xF{\xbL}-\xF{\xyL'}}. 
\end{align}
These quantities measure how much the fitness difference allocated to one of the three walk segments differs from what it would be assigned if $\xr$ and $\xs$ determined it exactly.
For example the $(\xxL,\xxL')$ edge is required to contain the fitness value $\xbet\xns\xr$.
Therefore the first walk segment can span a fitness distance of at most $\xbet\xns\xr$, but this happens exactly only if
$\xF{\xxL} - \xF{\xaL} = \xbet\xns\xr$ is chosen.
All other valid choices set the fitness value lower than this and $\epsilon_1$ measures the reduction of the segment's length.
As it will turn out only the point with all $\epsilon_i$ equal to zero contributes to the expectation value in leading order.
Intuitively any constant offset from the intended segment length corresponds to an effective reduction of $\xbet$ by a constant, resulting in an exponentially lower likelihood of walks being quasi-accessible.
Nonetheless we will carry the $\epsilon_i$ through the calculation.

The remaining inner expectation values are of the same form as the simple expectation of walks from $\xaL$ to $\xbL$ calculated in the previous section:
\begin{align}
	\label{simpleform}
	\E{\xwalksq{\xvL}{\xwL}|\xt}
	= \partial_\xt \e^{\xL\xGamL{\xvL}{\xwL}{\xt}}
	= \xgamL{\xvL}{\xwL}{\xt}\xL \e^{\xL\xGamL{\xvL}{\xwL}{\xt}}.
\end{align}
The expectation values are dominated by the exponential terms $\e^{\xL\xGamL{\xvL}{\xwL}{\xt}}$ with an additional linear factor $\xL$ resulting from the derivative.
However, the derivative also adds the term $\xgamL{\xvL}{\xwL}{\xt}$.
As an average over loci it can be seen that pointwise in $\xt$ and uniformly over $\xvL$ and $\xwL$, this quantity is bounded by a constant from above.
However the bound is not uniform in $\xt$.
At $\xt = 0$ it diverges, as can be seen from the expansion
\begin{align}
	\xL\xgamL{\xvL}{\xwL}{\xt} \sim \frac{\xdistL{\xvL}{\xwL}}{\xt}.
\end{align}
To avoid this issue we rewrite the expectation value including the sum resulting from application of the product rule of differentiation:
\begin{align}
    \label{sumform}
    \E{\xwalksq{\xvL}{\xwL}|\xt} =
    \sum_{\xl'=1}^\xL
    \xAed{\xv[\xl']}{\xw[\xl']}{\xt}
    \prod_{\xl\neq\xl'}\e^{\xGam{\xv[\xl]}{\xw[\xl]}{\xt}}.
\end{align}
In each summand the value is a product over terms, each of which depends only on quantities on a single locus and the bulk of the contributions of loci contribute simply the exponential $\e^{\xGam{\xv[\xl]}{\xw[\xl]}{\xt}} = \xAe{\xv[\xl]}{\xw[\xl]}{\xt}$.
Only the locus $l'$ gives a different contribution, namely the derivative of the exponential term, $\xAed{\xv[\xl']}{\xw[\xl']}{\xt}$.

Our goal is to bring eq. (\ref{ref1}) into the form of a sum over products, such that the product factorizes in the same sense as it does for a single expectation value.
In particular the current form is a sum of a product of three expectation values.
If we expand each expectation as shown in eq. (\ref{sumform}), we obtain three sums, each accounting for one special locus on which the corresponding derivative is taken.
We name these special loci $\xl_1$, $\xl_2$ and $\xl_3$, corresponding to the means in eq. (\ref{ref1}) in the order they appear there.
The sum in the middle term can be taken out of the $\p{\cdot}^{1-\alpha}$ form to give an upper bound, because $1-\alpha \in [0,1]$ and therefore the form is subadditive.
Having done so, the sum over the pair of edges on the genotype space may similarly be factorized over loci.
Each edge on the genotype graph corresponds to a step on one locus.
Therefore it is sufficient to sum over individual genotypes together with another special locus, and one edge on the allele graph corresponding to that locus.
We denote the sum over loci for these two edges $\xl_{12}$ and $\xl_{23}$ respectively.
The initial sum then factorizes over loci:
\begin{align}
    \Prob{\xwalksq{\xaL}{\xbL} \geq 1}
    \leq
    \sum_{\xl_1,\xl_2,\xl_3,\xl_{12},\xl_{23} = 1}^\xL
    \E{
        \prod_{\xl=1}^\xL \mathfrak{F}_{\xl}
    }.
\end{align}
Here $\mathfrak{F}_{\xl}$ is the resulting factor collecting all sums over quantities on locus $\xl$ and all factors of the product of the three expectations that are functions of quantities on locus $\xl$, as well as potentially e.g. a form $\sum_{\xx'}\xAp{\xx}{\xx'}{}$ if $\xl = \xl_{12}$.
$\mathfrak{F}_{\xl}$ is implicitly dependent on $\xl_1$, $\xl_2$, $\xl_3$, since these three variables decide whether the contribution resulting from any of the three expectation values has the usual exponential form or that of its derivative.
If $\xl \not\in \{\xl_1,\xl_2,\xl_3,\xl_{12},\xl_{23}\}$, then the contribution of all three expectation values and the edge sum is of the usual form, i.e. the exponential term of the expectation value and no sum over edges and we give it the name $\mathfrak{G}_\xl$:
\begin{align}
    \mathfrak{G}_\xl
    &= \sum_{\xx,\xy} \e^{
        \xGam{\xa}{\xx}{\xbet\xns\xr-\epsilon_{1}}
        +\p{1-\alpha}\xGam{\xx}{\xy}{\xbet\xs-\epsilon_{2}}
        +\xGam{\xy}{\xb}{\xbet\xns\xnr-\epsilon_{3}}
    }
    \\ &= \sum_{\xx,\xy}
        \xAe{\xa}{\xx}{\p{\xbet\xns\xr-\epsilon_{1}}}
        \p{\xAe{\xx}{\xy}{\p{\xbet\xs-\epsilon_{2}}}}^{1-\alpha}
        \xAe{\xy}{\xb}{\p{\xbet\xns\xnr-\epsilon_{3}}}.
\end{align}
If $\xl$ is equal to any of the set of special loci, then some of these exponential terms will be modified and there might be additional sums.
For example if $\xl$ is equal to $\xl_3$ and $\xl_{12}$, but not equal to any of the other special loci, then
\begin{align}
    \mathfrak{F}_\xl = \sum_{\xx,\xy,\xx'}
        \xAe{\xa}{\xx}{\p{\xbet\xns\xr-\epsilon_{1}}}
        \xAp{\xx}{\xx'}{}
        \p{\xAe{\xx'}{\xy}{\p{\xbet\xs-\epsilon_{2}}}}^{1-\alpha}
        \xAed{\xy}{\xb}{\p{\xbet\xns\xnr-\epsilon_{3}}}.
\end{align}
By assumption $\xs \neq 0$ and also $0 \neq \xr \neq 1$.
Then, with the fixed $\xs$ and $\xr$, $\mathfrak{G}_\xl$ is bounded away from zero everywhere except at the boundary, because as long as one of the matrix exponentials has non-zero argument, it contributes a finite term to the sum by adequate choice of $\xx$ and $\xy$ so that the indices of the matrix exponential become $(\xa,\xb)$.
More generally $\mathfrak{G}_l$ is also uniformly bounded over $\xs$ and $\xr$, since by definition of $\xs$ and $\xr$ at least one of the matrix exponential arguments must be at least $\frac{\xbet}{3}$, epsilon shifts notwithstanding.
This then allows us to write each $\mathfrak{F}_l$ as a product $\mathfrak{G}_l \mathfrak{H}_l$, with $\mathfrak{H}_l$ bounded away from infinity except at the mentioned boundary.
In the next section we will use the same approach with a more detailed handling of $\mathfrak{H}_l$, but here it is sufficient to apply such a simple uniform bound with a constant.

However first we consider the behavior at the boundary where all epsilon shifts force the matrix exponential arguments to become zero.
Due to the bounded degree of the graph, as $\xt \rightarrow 0$, the diagonal terms of the matrix exponential with argument $\xt$ drop to $1$ and the off-diagonal ones to $0$ uniformly.
If $\xa \neq \xb$, $\mathfrak{F}_\xl$ therefore falls to zero as all the $\epsilon_i$ reach their maximum boundary and similarly it falls to $1$ for $\xa = \xb$.
Since there is by assumption at least a finite fraction of loci with $\xa \neq \xb$, this then implies that eventually, at a finite distance to the boundary
\begin{align}
    \prod_{\xl=1}^\xL \mathfrak{F}_l \leq \O{C^\xL}
\end{align}
for some $C < 1$.
The special loci on which $\mathfrak{F}_l \neq \mathfrak{G}_l$ are not relevant to this, since there are only finitely many of them and each one is bounded.
The contribution to the probability from the boundary is therefore asymptotically zero, since the exponential decay in the integrand cannot be compensated by the additional $\xL^5$ factor from the special loci sum.

Returning to the general case away from the boundary, we can bound all $\mathfrak{H}_l$ with $\xl \in \{\xl_1,\xl_2,\xl_3,\xl_{12},\xl_{13}\}$ by some constant $C$ uniformly, yielding a factor of at most $C^5$, while all other $\mathfrak{H}_\xl$ are $1$.
This removes the dependence of the product on the particular choice of the special loci:
\begin{align}
    \Prob{\xwalksq{\xaL}{\xbL} \geq 1 } \leq \xL^5 C^5 \E{\prod_{\xl=1}^\xL \mathfrak{G}_\xl} + o(1)
    = \xL^5 C^5 \E{\e^{\xL\mathfrak{T}}}
\end{align}
with
\begin{align}
    \mathfrak{T} = \xmeanl{\ln\sum_{\xx,\xy}\e^{
        \xGam{\xa}{\xx}{\xbet\xns\xr-\epsilon_{1}}
        +\p{1-\alpha}\xGam{\xx}{\xy}{\xbet\xs-\epsilon_{2}}
        +\xGam{\xy}{\xb}{\xbet\xns\xnr-\epsilon_{3}}
    }}.
\end{align}
The $\epsilon_i$ are always non-negative in the valid domain and $\mathfrak{T}$ is decreasing in all of them.
Therefore we can give an upper bound by setting all of them to $0$ and obtain the upper bound on accessibility:
\begin{align}
    \Prob{\xwalksq{\xaL}{\xbL} \geq 1 }
    \leq \e^{\xL\p{\mathfrak{T}_0 + o(1)}}
\end{align}
with
\begin{align}
    \mathfrak{T}_0
    &=
    \xmeanls{\ln\sum_{\xx,\xy}\e^{
        \xGam{\xa}{\xx}{\xbet\xns\xr}
        +\p{1-\alpha}\xGam{\xx}{\xy}{\xbet\xs}
        +\xGam{\xy}{\xb}{\xbet\xns\xnr}
    }}
\end{align}
where it is assumed that $\xbet$ is constant in $\xL$.
This value is then independent of $\xL$ and if it is negative, the probability that $\xbL$ is accessible from $\xaL$ is asymptotically exponentially falling to zero.
We may choose $\alpha \in (0,1)$ as well as $\xs$ and $\xr$ freely except for their boundary values.
But specifically for $\alpha$ close to zero, we obtain the following expansion by matrix multiplication:
\begin{align}
    \mathfrak{T}_0 = \xmeanls{\xGam{\xa}{\xb}{\xbet}} - \alpha\xmeanls{\xmeanp{}{\xGam{\xx}{\xy}{\xbet\xs}}} + \O{\alpha^2}.
\end{align}
At $\xbets$ the zeroth order term is simply zero.
The coefficient of the linear order term is exactly $-\xmartFs{\xs}{\xr}{\xbet}$ and in the irregular type problem $\xr$ and $\xs$ can be chosen such that it is negative at $\xbets$.
With this choice there is then some suitable small $\alpha > 0$, so that $\mathfrak{T}_0$ is negative at $\xbets$.
$\mathfrak{T}_0$ is continuous as a function of $\xbet$ and therefore we can then also find some $\xbet > \xbets$ such that $\mathfrak{T}_0$ is still negative at the same choice of $\alpha$, $\xr$ and $\xs$.
This shows that the critical point $\xbet_c$ is strictly larger than $\xbets$ in the irregular case, if it exists at all.
\hfill $\Box$

\section{Proof of Theorem \ref{Thm1}.2.}

\newcommand{\tmpA}{\xwalksq{\xaL}{\xbL}}

\label{Sec:LowerBound}

In this section we derive a lower bound on accessibility, allowing us to show that the candidate threshold function in Theorem \ref{Thm1} does indeed satisfy the second side of the threshold requirement for accessibility setups of regular type.

\subsection{Moment bounds}

To prove the lower bound on (quasi-)accessibility, we use a generalization of the second moment method.
The idea of the second moment method is to bound the second moment of $\tmpA$ from above in order to apply the inequality
\cite{Alon2000,Hegarty2014}
\begin{align}
    \Prob{\tmpA > 0} \geq \frac{\E{\tmpA}^2}{\E{\tmpA^2}}.
\end{align}
$\tmpA$ is bounded from above through the maximum length of walks and the bounded degree limiting the possible choices in each step and therefore the second moment always exists.
In our proof method we do however find that, at least with our non-tight bounds on it, the second moment grows too quickly for some allele graphs to give a non-trivial bound.
On the other hand, for some class of allele graphs this bound may be used to obtain a sufficient bound.

To generalize the applicability of the result, we will use a modification of the second moment method which relies on a lower order moment.

\begin{lemma}
	Let $X$ be a random variable over the natural numbers (including zero) with finite moment $\E{X^{1+\xi'}}$ for some $\xi' > 0$, then
\begin{align}
	\Prob{X \geq 1} \geq \E{X}\e^{-\frac{\E{X\ln{X}}}{\E{X}}}
\end{align}
where an evaluation of $0 \ln 0$ is to be taken as $0$.
\end{lemma}
\begin{proof}

We know using H\"older's inequality that for all $0 < \xi \leq \xi'$:
\begin{align}
    \E{X}
    &= \E{ X \mathbb{I}_{X \geq 1} }
    \\ &\leq \E{X^{1+\xi}}^{\frac{1}{1+\xi}} \E{\mathbb{I}_{X\geq1}^{\frac{1}{1-\frac{1}{1+\xi}}}}^{1-\frac{1}{1+\xi}}
    \\ &= \E{X^{1+\xi}}^{\frac{1}{1+\xi}} \Prob{X\geq1}^{\frac{\xi}{1+\xi}}
\end{align}
and therefore
\begin{align}
    \Prob{X\geq1} \geq \p{ \frac{\E{X}^{1+\xi}}{\E{X^{1+\xi}}} }^{\frac{1}{\xi}}.
\end{align}
	Taking the limit of $\xi \rightarrow 0$ then completes the proof because
	\begin{align}
		\ln\E{X^{1+\xi}} = \ln\E{X} + \xi\frac{\E{X\ln X}}{\ln\E{X}} + \O{\xi^2}
	\end{align}
	where the value of $X\ln X$ is taken to be $0$ by analytic continuation. \qed
\end{proof}

Because the number of walks of length $\xN$ is at most exponential due to the bounded degree of $\xG$, while the probability of a walk to be quasi-accessible falls as fast as $\frac{1}{\xN!}$, the tail of $\tmpA$ is dominated by an exponential decay.
In particular all moments of $\tmpA$ exist.
This allows us to apply the lemma:
\begin{align}
    \Prob{\tmpA\geq1} \geq \E{\tmpA}\e^{-\mathfrak{K}}
\end{align}
where
\begin{align}
    \mathfrak{K} = \E{\frac{\tmpA}{\E{\tmpA}}\ln\tmpA}.
\end{align}
In order to prove Theorem \ref{Thm1}.2 we need to show that $\liminf \Prob{\xwalksq{\xaL}{\xbL} \geq 1} > 0$ with the proposed threshold function. In particular we will choose for this section $\xbet = c_L+\frac{\eta}{\xL}$ with some constant $\eta$.

Per Corollary \ref{CorA} then $\E{\tmpA}$ converges to a non-zero value.
This assures that it is sufficient to show that $\mathfrak{K}$ does not diverge.
The following method of bounding $\mathfrak{K}$ adapts the idea used in \cite{Martinsson2018} to account for the correlations of accessible walks using the notion of shortcuts or arcs to obtain alternative walks from a focal one.

Let $X_\pi$ be the indicator variable that the walk $\pi$ is quasi-accessible, then
\begin{align}
    \mathfrak{K} = \sum_{\pi}\E{\frac{X_{\pi}}{\E{\tmpA}} \ln\tmpA}
\end{align}
where the sum is over all walks from $\xaL$ to $\xbL$ on $\xGL$ or equivalently all valid walks on $\xGL'$.

\begin{figure}
    \begin{center}
	\includegraphics{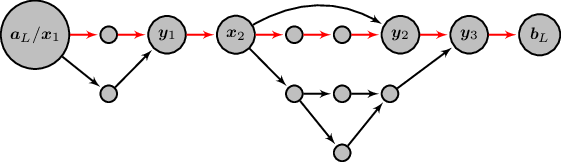}
    \end{center}
    \caption{
    	\label{Fig7}
        Illustration of the estimation of $\xwalksq{\xaL}{\xbL}$.
        The short horizontal arrows (red) indicate the focal walk $\pi$ from $\xaL$ to $\xbL$.
        Relative to $\pi$ there is one non-trivial arc from $\xxL_1$ to $\xyL_1$, one non-trivial arc from $\xxL_2$ to $\xyL_2$ and two non-trivial arcs from $\xxL_2$ to $\xyL_3$.
        Between any two adjacent vertices on $\pi$ there is furthermore one trivial arc.
        All other non-empty path subgraphs are not arcs, since they intersect $\pi$ more than twice.
        Each walk is fully specified by choice of one of the arcs for each pair of loci.
        In fact it would be sufficient to choose one out-going arc per site to account for all walks.
        By the construction of quasi-accessibility the graph is guaranteed to be free of cycles as in the example (color figure online)
    }
\end{figure}

Similarly we can expand the right hand $\tmpA$ over individual walks.
A graphical example for the situation described in the following can be seen in Fig.~\ref{Fig7}.
We will however intentionally over-count these in the following way:
Each valid walk $\pi'$ from $(\xaL,1)$ to $(\xbL,1)$ in $\xGL'$ trivially crosses $\pi$ in at least two vertices, namely $(\xaL,1)$ and $(\xbL,1)$.
Furthermore if we list out for each valid walk $\pi'$ the vertices it shares with $\pi$ in $\xGL'$, then the segment of $\pi'$ between two adjacent vertices $\xxL \in \xVL'$ and $\xyL \in \xVL'$ in that list does not intersect $\pi$ a third time in $\xGL'$.
We call such a segment on $\xGL'$ an \emph{arc} through $\xxL$ and $\xyL$ on $\pi$.
An arc is said to be \emph{trivial} if it is a segment of $\pi$ itself.
Immediately from the definition a trivial arc can only contain a single edge.
We denote the number of non-trivial arcs which are accessible by $\xwalksqrestrict{\xxL}{\xyL\pi}$.
Each walk $\pi'$ generates at most one arc through $\xxL$ and $\xyL$ on $\pi$.
Also each walk $\pi'$ is uniquely identified by the set of arcs it generates on $\pi$ and $\pi'$ is accessible on $\xGL'$ if and only if all of the arcs it generates on $\pi$ are accessible.
Therefore we can bound for any valid walk $\pi$:
\begin{align}
    \xwalksq{\xaL}{\xbL} \leq \prod_{\xxL,\xyL\in\xVL'}\p{1+\xwalksqrestrict{\xxL}{\xyL\pi}}.
\end{align}
With this we have
\begin{align}
    \mathfrak{K} \leq \sum_{\xxL,\xyL\in\xVL'}\sum_{\pi}I_{\xxL\xyL\pi}\E{\frac{X_\pi}{\E{\tmpA}} \ln\p{1+\xwalksqrestrict{\xxL}{\xyL\pi}}}
\end{align}
where $I_{\xxL\xyL\pi}$ is an indicator variable which is $1$ iff $\pi$ contains $\xxL$ and $\xyL$ in $\xGL'$ and $0$ otherwise.

Conditioned on the two fitness values $\xF{\xxL}$ and $\xF{\xyL}$, $\xwalksqrestrict{\xxL}{\xyL\pi}$ becomes independent of $X_\pi$ since $\xxL$ and $\xyL$ are the only vertices whose fitness values influence both the quasi-accessibility of candidate arcs and $\pi$:
\begin{align}
    \mathfrak{K} \leq \sum_{\xxL,\xyL\in\xVL'}\sum_{\pi}I_{\xxL\xyL\pi}\E{\E{\frac{X_\pi}{\E{\tmpA}}|\xF{\xxL},\xF{\xyL}} \E{\ln\p{1+\xwalksqrestrict{\xxL}{\xyL\pi}}|\xF{\xxL},\xF{\xyL}}}.
\end{align}
For convenience we also assume that the conditioning of the fitness values $\xF{(\xaL,1)}$ and $\xF{(\xbL,1)}$ to a difference of $\xbet$ is contained in the outer expectation.

Currently $\xwalksqrestrict{\xxL}{\xyL\pi}$ is stochastically independent of $X_\pi$, but still explicitly dependent on $\pi$ in the choice of candidate arcs that need to be counted.
We can remove this dependence by loosening the restriction that included arcs must not be trivial and must not intersect $\pi$ except at $\xxL$ and $\xyL$.
Doing so $\xwalksqrestrict{\xxL}{\xyL\pi}$ is upper bounded by $\xwalksqrestrict{\xxL}{\xyL}$, where the lack of third index indicates the loosened restriction.
The resulting bound is not in general good enough for all choices of $\xxL$ and $\xyL$ in the sum.
We will later revisit and adjust it for these special cases.

Because the logarithm is concave, the mean over it can be bounded by exchange of the two.
Let $\downarrow\xxL$ be the projection of $\xxL\in\xGL'$ on the first component or equivalently the $1$-section of $\xGL'$.
Compared to all walks on $\xGL'$ generated from walks on $\xGL$ from $\downarrow\xxL$ to $\downarrow\xyL$, arcs from $\xxL$ to $\xyL$ in $\xGL'$ are more restricted in the number of times vertices with projection $\downarrow\xxL$ or $\downarrow\xyL$ may or must be visited.
Therefore the expectation over $\xwalksqrestrict{\xxL}{\xyL}$ may be bounded by the expectation over $\xwalksq{\downarrow\xxL}{\downarrow\xyL}$.
\begin{align}
    \mathfrak{K} \leq \E{\sum_{\xxL,\xyL\in\xVL'}\E{\frac{\sum_{\pi}I_{\xxL\xyL\pi}X_\pi}{\E{\tmpA}}|\xF{\xxL},\xF{\xyL}} \ln\E{1+\xwalksq{\downarrow\xxL}{\downarrow\xyL}|\xF{(\downarrow\xxL,1)},\xF{(\downarrow\xyL,1)}}}.
\end{align}

\begin{figure}
    \begin{center}
	\includegraphics{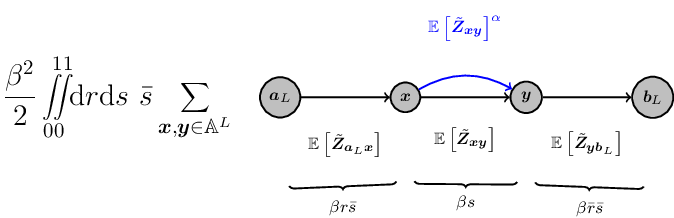}
    \end{center}
    \caption{
    \label{Fig8}
        Graphical representation of the bound on $\mathfrak{K}$ given in Eq.~(\ref{Keq}).
        The fitness range $\xbet$ is split into three segments corresponding to paths $\xaL \rightarrow \xxL$, $\xxL \rightarrow \xyL$ and $\xyL \rightarrow \xbL$, indicated by straight arrows (black).
        Each segment contributes the given expectation value conditioned on the specified fitness difference between its endpoints into a product.
        The curved arrow (blue) represents the contribution of all arcs from $\xxL$ to $\xyL$, which contribute the given $\alpha$-dependent factor (color figure online)
    }
\end{figure}

Similarly all walks $\pi$ through $\xxL$ and $\xyL$ can be separated into three segments from $(\xaL,1)$ to $\xxL$, from $\xxL$ to $\xyL$ and from $\xyL$ to $(\xbL,1)$.
Each walk is uniquely determined by these three segments and for any choice of these segments forming a valid walk, their accessibility is independent under the conditioning since valid walks are selfavoiding in $\xGL'$.
Taking all triples of walk segments from $\xaL$ to $\downarrow\xxL$, from $\downarrow\xxL$ to $\downarrow\xyL$ and from $\downarrow\xyL$ to $\xbL$, all valid walks from $\xaL$ to $\xbL$ through any copy of the genotypes $\downarrow\xxL$ and $\downarrow\xyL$ in $\xGL'$ are generated.
This allows together with the previous arguments for the bound
\begin{align}
    \mathfrak{K} \leq \E{\sum_{\xxL,\xyL\in\xVL}\frac{
    \E{\xwalksq{\xaL}{\xxL} | \xF{\xxL},\xF{\xyL}}
    \E{\xwalksq{\xxL}{\xyL} | \xF{\xxL},\xF{\xyL}}
    \E{\xwalksq{\xyL}{\xbL} | \xF{\xxL},\xF{\xyL}}
}{\E{\tmpA}} \ln\p{1+\E{\xwalksq{\xxL}{\xyL}|\xF{\xxL},\xF{\xyL}}}}.
\end{align}
Further we use that the logarithm can be bounded from above by any power law $\ln(x) \leq \bar\alpha (x-1)^\alpha$ for $x \geq 1$, $0 < \alpha < 1$ and a constant $\bar\alpha$ depending on $\alpha$.
In particular $\bar\alpha$ as a function of $\alpha$ can be chosen so that it is bounded except around $\alpha = 0$, where it must diverge.
Therefore, as long as we choose later any non-zero but constant $\alpha$, the additional factor $\bar\alpha$ will not change the asymptotic order of $\mathfrak{K}$.
All in all we have the following bound, which is also represented graphically in Fig.~\ref{Fig8}:
\begin{align}
   \label{Keq}
    \mathfrak{K} \leq \E{\bar\alpha\sum_{\xxL,\xyL\in\xVL}\frac{
    \E{\xwalksq{\xaL}{\xxL} | \xF{\xxL},\xF{\xyL}}
    \E{\xwalksq{\xxL}{\xyL} | \xF{\xxL},\xF{\xyL}}^{1+\alpha}
    \E{\xwalksq{\xyL}{\xbL} | \xF{\xxL},\xF{\xyL}}
}{\E{\tmpA}}}.
\end{align}

The remaining expectation values depend only on the differences of the fitness values that they are conditioned on, not the actual placement of that difference.
It is therefore convenient to use the variables $\xs$ and $\xr$ with $\xns = 1-\xs$ and $\xnr = 1-\xr$ introduced previously,
such that
\begin{align}
    \xF{\xxL}-\xF{\xaL} &= \xns\xr\xbet, \\
    \xF{\xyL}-\xF{\xxL} &= \xs\xbet, \\
    \xF{\xbL}-\xF{\xyL} &= \xns\xnr\xbet,
\end{align}
with which the outer expectation value of $\mathfrak{K}$ can be expressed as an integral over the unit square $\p{\xs,\xr} \in [0,1]^2$ with a surface element $\frac{\xbet^2\xns}{2}\mathrm{d}r\mathrm{d}s$, which through the factor $\frac{1}{2}$ already conditions on $\xF{\xxL}$ and $\xF{\xyL}$ being correctly ordered.
We need to show that this integral is asymptotically bounded by a constant in order to show that $\mathfrak{K}$ is asymptotically bounded by a constant from above as we intend.

The expectation value $\E{\tmpA}$ may be bounded using eq. (\ref{simpleform}):
\begin{align}
    \E{\xwalksq{\xaL}{\xbL}}
    = \Theta(1) \xL \e^{\xL\xGamL{\xaL}{\xbL}{\xbet}}.
\end{align}
The same bound does not in general apply to the other expectation values uniformly over the integration domain due to the divergence of the constant term with vanishing fitness difference.
For this reason, we split the integration region.
For some sufficiently small constant $\epsilon > 0$ we will consider integration in the regions with $\xs \in [0,\epsilon]$ and $\xs \in [\epsilon, 1]$ separately and name the corresponding contributions to $\mathfrak{K}$ accordingly with an index.

\subsection{Case $s \in [\epsilon,1]$}

In the interval $[\epsilon,1]$, $\xs$ is bounded away from zero and therefore using eq. (\ref{simpleform}), the expectation values $\E{\xwalksq{\xxL}{\xyL} | \xbet\xs}$ can be bounded uniformly by
\begin{align}
    \E{\xwalksq{\xxL}{\xyL} | \xbet\xs}
	\leq \O{1}\xL \e^{\xL\xGamL{\xxL}{\xyL}{\xbet\xs}}.
\end{align}
For the remaining expectation values we follow the procedure used in Sect.~\ref{Sec:ImprovedBound} and expand with eq. (\ref{sumform}) to obtain a sum of locus-factorized terms.
We name the special loci according to the walk segment's index.
As we have already expanded the contributions for the second segment, only the first and third remain:
\begin{align}
    \mathfrak{K}_{[\epsilon,1]}
    = \O{1}\bar\alpha\xL^\alpha\sum_{\xl_1,\xl_3=1}^\xL \E{\prod_{\xl=1}^\xL \mathfrak{F}_l}
    = \O{1}\bar\alpha\xL^\alpha\sum_{\xl_1,\xl_3=1}^\xL \E{\prod_{\xl=1}^\xL \mathfrak{G}_l\mathfrak{H}_l}.
\end{align}
Again, the usual form for loci $\xl \not\in \{l_1,l_3\}$ can be given through the exponential terms in the expectation values
\begin{align}
    \mathfrak{G}_l = \sum_{\xx,\xy\in\xV}\e^{\xGam{\xa}{\xx}{\xbet\xns\xr} + \p{1+\alpha}\xGam{\xx}{\xy}{\xbet\xs} + \xGam{\xy}{\xb}{\xbet\xns\xnr} - \xGam{\xa}{\xb}{\xbet}}
\end{align}
and again for loci $\xl \in \{l_1,l_3\}$ one or more of the exponential factors will be replaced by their derivatives.
For the same reasons as used previously, in these cases $\mathfrak{H}_l$ is uniformly bounded by a constant and therefore
\begin{align}
    \mathfrak{K}_{[\epsilon,1]} &= \O{1} \bar\alpha\xL^{2+\alpha} \e^{\xL\mathfrak{T}}
\end{align}
where
\begin{align}
    \mathfrak{T} = 
    \xmeanl{\ln\sum_{\xx,\xy\in\xV}\e^{
        \xGam{\xa}{\xx}{\xbet\xns\xr}
        +\p{1+\alpha}\xGam{\xx}{\xy}{\xbet\xs}
        +\xGam{\xy}{\xb}{\xbet\xns\xnr}
        -\xGam{\xa}{\xb}{\xbet}
    }}.
\end{align}
$\mathfrak{T}$ can be considered a function of $\xbet$, $\xs$, $\xr$ and $\alpha$.
At $\alpha = 0$, it is always $0$ as can be verified by matrix multiplication.
The first derivative towards $\alpha$ at $\alpha=0$ is found to be exactly $\xmartF{\xs}{\xr}{\xbet}$.
It is therefore possible to bound
\begin{align}
    \mathfrak{T} = \alpha\xmartF{\xs}{\xr}{\xbet} + \O{\alpha^2}.
\end{align}
In the regular case, as $\xbet$ converges to $\xbetb$, $\xmartF{\xs}{\xr}{\xbet}$ is eventually bounded from above by a negative constant in the region $\xs \in [\epsilon,1-\epsilon]$, so that in this region the integrand falls exponentially quickly to zero for suitable choice of $\alpha > 0$, resulting in no asymptotic contribution to $\mathfrak{K}$.
In the interval $\xs \in [1-\epsilon,1]$ we need to account for the boundary term at $\xs = 1$.
At $\xs = 1$, $\mathfrak{T}$ is exactly $\alpha\xGamL{\xaL}{\xbL}{\xbet}$.
At the candidate threshold function $\xGamL{\xaL}{\xbL}{\xbet}$ simply evaluates to $-\frac{\ln\xL}{\xL}$ up to irrelevant higher orders in $\xL$.
By assumptions for the regular case we also have that the derivative $\partial_\alpha\partial_\xs\mathfrak{T}$ is positive at $(\xs,\alpha,\xbet) = (1,0,\xbetb)$, so that
\begin{align}
    \mathfrak{T} \leq -\alpha\frac{\ln\xL}{\xL} + \O{\p{\frac{\ln\xL}{\xL}}^2} - c\xns\alpha
\end{align}
for some $c > 0$.
The term $-\alpha\frac{\ln\xL}{\xL}$ exactly compensates a factor $\xL^\alpha$ to the integrand of $\mathfrak{K}$ and with a factor $\xns$ in the surface element of the integration, the contribution to $\mathfrak{K}$ from $\xs \in [1-\epsilon,1]$ is then for suitably small constant $\alpha > 0$:
\begin{align}
    \mathfrak{K}_{[1-\epsilon,1]} = \O{1}\xL^2\intl{\xs}{1-\epsilon}{1} \xns\e^{-c\xL\xns\alpha} = \O{1}.
\end{align}

\subsection{Case $\xs \in [0,\epsilon]$}
For the integration interval $[0,\epsilon]$ we will fix $\alpha = 1$ and since we cannot apply the simple bound to the expectation $\E{\xwalksq{\xxL}{\xyL}|{\xbet\xs}}^{1+\alpha}$ used before uniformly in this region, we will expand it using the sum form of the expectation value.
Since $1+\alpha = 2$ now, there will effectively be two additional sums resulting from this, for which we label the corresponding locus variables $\xl_{21}$ and $\xl_{22}$.
Again, we bring the contribution into the form
\begin{align}
    \mathfrak{K}_{[0,\epsilon]}
    = \O{1}\xL^{-1}\sum_{\xl_1,\xl_{21},\xl_{22},\xl_{3}} \E{\mathfrak{F}_l}
    = \O{1}\xL^{-1}\sum_{\xl_1,\xl_{21},\xl_{22},\xl_{3}} \E{\mathfrak{G}_l\mathfrak{H}_l}.
\end{align}
Here, since all expectation values in the numerator of $\mathfrak{K}$ were expanded into sums, only a single factor $\xL^{-1}$ remains from the expectation value in its denominator.
The usual form $\mathfrak{G}_l$ is unchanged from the region $[\epsilon,1]$ except for the choice $\alpha = 1$.
As before $\mathfrak{H}_l$ can be bounded by a constant for all special $\xl$, but this will turn out not to be sufficient here.
Suppose we used such a bound, then we would obtain
\begin{align}
    \mathfrak{K}_{[0,\epsilon]}
    \leq \O{1} \xL^3 \E{\e^{\xL\mathfrak{T}}}
\end{align}
where $\mathfrak{T}$ is unchanged from the previous integration region except for the choice $\alpha = 1$.
At $\xs = 0$ only terms with $\xx = \xy$ can contribute to $\mathfrak{T}$ and so it becomes $0$ by matrix multiplication.
The first derivative towards $\xs$ can be formed directly, using that derivatives of matrix exponentials correspond to multiplication with the matrix exponent. Using that $\xAp{\xx}{\xy}{}\mathbb{I}_{\xx\xy} = 0$ since the allele graph is simple, the derivative evaluates exactly to $-\xbet\xgamL{\xaL}{\xbL}{\xbet}$, so that:
\begin{align}
    \mathfrak{T} = -\xbet\xgamL{\xaL}{\xbL}{\xbet}\xs + \O{\xs^2}.
\end{align}
As $\xbet\xgamL{\xaL}{\xbL}{\xbet}$ is strictly positive and bounded away from zero asymptotically, this shows that $\epsilon$ can always be chosen such that $\mathfrak{T}$ is negative for $\xs \in (0,\epsilon]$ with negative first derivative at $\xs = 0$.
Consequently the integration at the boundary is of the form
\begin{align}
    \mathfrak{K}_{[0,\epsilon]} \leq \O{1} \xL^3 \intl{\xs}{0}{\epsilon} \e^{-c\xL\xs} = \O{\xL^2}
\end{align}
for some constant $c > 0$.
In contrast to the boundary at $\xs = 1$ the surface element does not contribute here and does not yield an additional factor $\xL^{-1}$.
The naive bound shown above is not sufficient and two powers of $\xL$ remain that we have to suppress.

To cancel these factors, we need to bound the terms $\mathfrak{H}_l$ more carefully around small $\xs$ instead of applying uniform constant bounds.
In particular it would be sufficient to show that these terms introduce at least two factors $\xs$ into the integrand, since the integration over $\xs^n \e^{-c\xL\xs}$ would result in a value of order $\xL^{-1-n}$ instead of just $\xL^{-1}$.
Depending on the choices of distances between $\xxL$ and $\xyL$ and the choices of the special loci it is possible to provide these two factors.
However not all combinations of these choices yield such a factor.
The problematic cases will however turn out to be marginal in the sense that they only apply to a fraction $\frac{1}{\xL}$ or $\frac{1}{\xL^2}$ of the summands in the sums over special loci.
Each such factor $\xL^{-1}$ offsets the need for one $\xs$ factor in the integrand, allowing the total contribution to $\mathfrak{K}$ to still be constant.
In the following we list all of the relevant combinations and show their contributions of $\xs$ orders.

The method of bounding $\mathfrak{H}_l$ is to consider the small-$\xs$ behavior of the factors $\xAe{\xx}{\xy}{\xs\xbet}$ and $\xAed{\xx}{\xy}{\xs\xbet}$ which appear in it.
In particular, depending on the distance we have
\begin{align}
    \label{expansion}
    \xAe{\xx}{\xy}{\xs\xbet} &=
    \begin{cases}
        1 + \O{\xs^2}   & \xx = \xy \\
        \O{\xs^{\xdist{\xx}{\xy}}}         & \xx \neq \xy
    \end{cases} \\
    \xAed{\xx}{\xy}{\xs\xbet} &=
    \begin{cases}
        \O{\xs}   & \xx = \xy \\
        \O{\xs^{\xdist{\xx}{\xy}-1}}         & \xx \neq \xy
    \end{cases}
\end{align}
and due to the bounded degree of the graph, all of these bounds are uniform.

Using the bounds above, we can obtain the necessary factors of $\xs$.
First consider the case of all four special loci distinct.
We then have for the two loci $\xl_{21}$ and $\xl_{22}$:
\begin{align}
    \mathfrak{H}_{\xl_{2i}}
    &= \frac{
        \sum_{\xx[\xl_{2i}],\xy[\xl_{2i}]\in\xV}
            \xAe{\xa[\xl_{2i}]}{\xx[\xl_{2i}]}{\xbet\xns\xr}
            \xAe{\xx[\xl_{2i}]}{\xy[\xl_{2i}]}{\xbet\xs}
            \xAed{\xx[\xl_{2i}]}{\xy[\xl_{2i}]}{\xbet\xs}
            \xAe{\xy[\xl_{2i}]}{\xb[\xl_{2i}]}{\xbet\xns\xnr}
        }{
            \sum_{\xx[\xl_{2i}],\xy[\xl_{2i}]\in\xV}
            \xAe{\xa[\xl_{2i}]}{\xx[\xl_{2i}]}{\xbet\xns\xr}
            [\xAe{\xx[\xl_{2i}]}{\xy[\xl_{2i}]}{\xbet\xs}]^2
            \xAe{\xy[\xl_{2i}]}{\xb[\xl_{2i}]}{\xbet\xns\xnr}
        }.
\end{align}
From eq. (\ref{expansion}) we can see that for all distances
\begin{align}
    \xAe{\xx}{\xy}{\xbet\xs} \xAed{\xx}{\xy}{\xbet\xs} = \O{\xs}
\end{align}
and therefore each of $\mathfrak{H}_{\xl_{21}}$ and $\mathfrak{H}_{\xl_{22}}$ contribute at least one factor $\xs$, resulting in a sufficient contribution of $\xs^2$ as explained above.
If not all of the four loci are distinct the form of $\mathfrak{H}_{2i}$ will be different.
However, the only modifications are in the placement of derivatives of matrix exponentials.
As long as still $\xl_{21} \neq \xl_{22}$, the relevant terms which are small around $\xs = 0$, namely the exponentials for the second walk segment, remain unchanged.

Therefore the remaining cases are for $\xl_{21} = \xl_{22}$, for which we will write $\xl_2$.
This equality reduces the number of summands to consider by a factor $\xL^{-1}$ as discussed before and consequently we need to find only one factor $\xs$.
In particular if either $\xl_1$ or $\xl_3$ are equal to $\xl_2$ as well, then the weight of these cases is reduced by another factor $\xL^{-1}$, so that no $\xs$ is required anymore.
Therefore, we can focus only on the case where $\xl_1$, $\xl_2$ and $\xl_3$ are all distinct.
For this case the contribution of locus $\xl_2$ is
\begin{align}
    \mathfrak{H}_{\xl_2}
    = &\frac{
        \sum_{\xx[\xl_2],\xy[\xl_2]\in\xV}
            \xAe{\xa[\xl_2]}{\xx[\xl_2]}{\xbet\xns\xr}
            [\xAed{\xx[\xl_2]}{\xy[\xl_2]}{\xbet\xs}]^2
            \xAe{\xy[\xl_2]}{\xb[\xl_2]}{\xbet\xns\xnr}
        }{
        \sum_{\xx[\xl_2],\xy[\xl_2]\in\xV}
            \xAe{\xa[\xl_2]}{\xx[\xl_2]}{\xbet\xns\xr}
            [\xAe{\xx[\xl_2]}{\xy[\xl_2]}{\xbet\xs}]^2
            \xAe{\xy[\xl_2]}{\xb[\xl_2]}{\xbet\xns\xnr}
        }.
\end{align}
Following again eq. (\ref{expansion}), the numerator is of order $\O{\xs^2}$ except if $\xdist{\xx[\xl_2]}{\xy[\xl_2]} = 1$, in which case there is a zeroth order contribution.
The latter case requires additional considerations to resolve.

First, we consider the subcase with $\xdistL{\xxL}{\xyL} \geq 2$.
In this case it is possible that $\xdist{\xx[\xl_2]}{\xy[\xl_2]} = 1$, but if this is the case we always have another locus $\xl'$ with $\xdist{\xx[\xl']}{\xy[\xl']} \geq 1$.
Following the separation of edges in the previous section, we can handle one such locus as a special locus in exchange for another sum of order $\xL$.
However, the $\xx[\xl'] \rightarrow \xy[\xl']$ factor contributions in $\mathfrak{H}_{l'}$'s numerator will then always be $\left[\xAe{\xx[\xl']}{\xy[\xl']}{\xbet\xs}\right]^2$ without any derivatives since $l' \neq \xl_2$.
From eq. (\ref{expansion}), such a factor results in a factor $\xs^2$ compensating the additional $\xL$ sum as well as the required $\xs$ factor to the integrand.

\begin{figure}
    \begin{center}
	\includegraphics{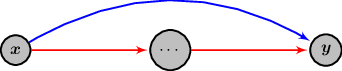}
    \end{center}
    \caption{
        Distance-$1$ case for $\xwalksqrestrict{\xxL}{\xyL\pi}$.
        The dots represent an arbitrary acyclic subgraph.
        The distance-1 path indicated by the curved arrow (blue) is always accessible, assuming that the initial and final fitness values for $\xxL$ and $\xyL$ are correctly ordered, which we enforce through the integral boundaries.
        The focal path is also conditioned on being accessible.
        Assuming that $\pi$ is the focal path indicated by the straight arrows (red) and $\pi'$ the distance-1 path (blue), then in $\xwalksqrestrict{\xaL}{\xbL\pi'}$ all paths except $\pi'$ are counted, while in $\xwalksqrestrict{\xaL}{\xbL\pi}$ at least one accessible path ($\pi$) is excluded, but more paths that are not arcs may also be excluded.
        Therefore $\xwalksqrestrict{\xaL}{\xbL\pi} \leq \xwalksqrestrict{\xaL}{\xbL\pi'}$ under the stated conditioning
        (color figure online)
        \label{Fig:distance1}
    }
\end{figure}

The only remaining case is then $\xdistL{\xxL}{\xyL} = 1$.
For this case the contribution to $\mathfrak{K}$ is indeed not bounded as we require.
However, this contribution turns out to be an overcounting issue introduced by our loosening of the arc restrictions on $\xwalksqrestrict{\xxL}{\xyL\pi}$.
Specifically, if $\xdistL{\xxL}{\xyL} = 1$, we will enforce the restriction that $\xwalksqrestrict{\xxL}{\xyL\pi}$ should not count the direct walk segment $\xxL \rightarrow \xyL$ if $\pi$ is taking this direct step.
Since the direct step is always accessible given that $\xF{\xxL}$ and $\xF{\xyL}$ are ordered correctly, this segment contributes exactly $1$ to the expectation value $\E{\xwalksq{\xxL}{\xyL}|\xbet\xs}$, which we can therefore substract from it.
This is possible even if $\pi$ does not use this direct step since $\xwalksqrestrict{\xxL}{\xyL\pi} \leq \xwalksqrestrict{\xxL}{\xyL\pi'}$ if $\pi$ does not use the trivial arc, but $\pi'$ does (Figure \ref{Fig:distance1}).
With this modification the value of $\mathfrak{H}_{l_2}$ becomes
\begin{align}
    \mathfrak{H}_{\xl_2}
    = &\frac{
        \sum_{\xx[\xl_2],\xy[\xl_2]\in\xV}
            \xAe{\xa[\xl_2]}{\xx[\xl_2]}{\xbet\xns\xr}
            \xAed{\xx[\xl_2]}{\xy[\xl_2]}{\xbet\xs}\p{\xAed{\xx[\xl_2]}{\xy[\xl_2]}{\xbet\xs}-1}
            \xAe{\xy[\xl_2]}{\xb[\xl_2]}{\xbet\xns\xnr}
        }{
        \sum_{\xx[\xl_2],\xy[\xl_2]\in\xV}
            \xAe{\xa[\xl_2]}{\xx[\xl_2]}{\xbet\xns\xr}
           [\xAe{\xx[\xl_2]}{\xy[\xl_2]}{\xbet\xs}]^2
            \xAe{\xy[\xl_2]}{\xb[\xl_2]}{\xbet\xns\xnr}
        }.
\end{align}
Since $\xAp{\xx[\xl_2]}{\xy[\xl_2]}{} = 1$, the leading order in the numerator is now $\O{\xs}$, which is sufficient to obtain a bounded contribution to $\mathfrak{K}$.

All in all, the total contributions to $\mathfrak{K}$ are bounded in $\xL$ at the candidate threshold function for the regular type, implying that there is a constant $C > 0$, such that
\begin{align}
    \liminf \Prob{\xwalksq{\xaL}{\xbL} \geq 1} \geq C.
\end{align}
This completes the proof of Theorem \ref{Thm1}.2, showing that for
regular setups there is indeed a threshold function $c_L$ at which
accessibility jumps in a $\frac{1}{\xL}$ window from $0$ to a non-zero
value of at least $C$. \hfill $\Box$

It remains to improve this bound from non-zero $C$ to $C=1$, which we expect can be done as mentioned in Section \ref{Sec:Statements},
Remark \ref{remark:weak}.

\section{Discussion}

Cartesian power graphs provide a natural framework for describing
genotype spaces composed of sequences of elements drawn from a
finite set of alleles $\xV$. The allele graph $\xG$ encodes the possible
mutational transitions on this set. Once the genotype-fitness map is
specified according to the HoC model, which assigns fitness values to
genotypes as i.i.d. continuous random variables, the rank order properties of the
resulting fitness landscape are uniquely determined by $\xG$.

Here we have focused on the existence of fitness-monotonic paths as a
measure of evolutionary accessibility \cite{Franke2011}, and proved precise results for
the critical fitness difference quantile $\xbetc$ above which accessible paths
exist with positive probability. Our results quantify how
accessibility increases with an increase of the number of alleles $\xall$ \cite{Wu2016,Zagorski2016}, and
decreases when mutational transitions are blocked or become
unidirectional. For certain allele graphs, such as the path graph over
three or more alleles, accessible paths do not exist for any fitness
difference. Moreover, a criterion based on the behavior of
Martinsson's function $\xmartF{\xs}{\xr}{\xbet}$ identifies allele graphs for which the behavior
of the expected number of accessible paths is not informative about
the existence of paths. In the words of Berestycki et al. \cite{Berestycki2014}, for such
allele graphs the expectation does not \textit{``tell the truth''}.

The HoC accessibility problem considered here is conceptually
appealing, because under the assumption of i.i.d. random fitness
values, landscape accessibility is determined solely by the structure of the
genotype space. However, the HoC model is not biologically realistic,
as empirical fitness landscapes display varying degrees of fitness
correlations \cite{Szendro2013,deVisser2014}. The accessibility properties of correlated fitness
landscapes differ significantly from those of the HoC model \cite{Krug2019}. For
example, for the much studied class of NK fitness landscapes,
accessibility is determined by the structure of the interaction graph,
and is low for most common
structures \cite{Hwang2018,Schmiegelt2014}. Previous work on accessibility of NK fitness landscapes has been restricted to the biallelic case, and exploring
the interplay between the allele graph and the interaction graph in
determining evolutionary accessibility constitutes an interesting
problem for future research.

\section*{Acknowledgements} This work was supported by DFG within SPP
1590 \textit{Probabilistic structures in evolution}. We thank David
Augustin for his contributions to the early stages of this project, and Alexander Drewitz and Muhittin Mungan for discussions.


\begin{thebibliography}{100}
\providecommand{\url}[1]{{#1}}
\providecommand{\urlprefix}{URL }
\expandafter\ifx\csname urlstyle\endcsname\relax
  \providecommand{\doi}[1]{DOI~\discretionary{}{}{}#1}\else
  \providecommand{\doi}{DOI~\discretionary{}{}{}\begingroup
  \urlstyle{rm}\Url}\fi

\bibitem{Alon2000} Alon, N., Spencer, J.: {The Probabilistic Method}.
  \newblock 2nd ed., Wiley, New York (2000).
  
\bibitem{Altenberg2015}
Altenberg, L.: {Fundamental properties of the evolution of mutational robustness}.
\newblock Preprint arXiv:1508.07866 (2015)

\bibitem{Berestycki2016}
Berestycki, J., Brunet, {\'E}., Shi, Z.: {The number of accessible paths in the
  hypercube}.
\newblock Bernoulli \textbf{22}, 653--680 (2016)

\bibitem{Berestycki2014}
Berestycki, J., Brunet, {\'E}., Shi, Z.: {Accessibility percolation with
  backsteps}.
\newblock ALEA, Lat. Am. J. Probab. Math. Stat. \textbf{14}, 45--62 (2017)

\bibitem{Carneiro2010}
Carneiro, M., Hartl, D.L.: {Adaptive landscapes and protein evolution}.
\newblock {Proc. Nat. Acad. Sci. USA} \textbf{107}, 1747--1751 (2010)

\bibitem{Crona2013}
Crona, K., Greene, D., Barlow, M.: {The peaks and geometry of fitness
  landscapes}.
\newblock {J. Theor. Biol.} \textbf{318}, 1--10 (2013)

\bibitem{Fragata2019} Fragata, I., Blanckaert, A., Louro, M.A.D., Liberles, D.A., Bank, C.:
  {Evolution in the light of fitness landscape theory}.
  \newblock {Trends in Ecology \& Evolution} \textbf{34}, 69--82 (2019)

\bibitem{Franke2011}
Franke, J., Kl{\"o}zer, A., de~Visser, J.A.G.M., Krug, J.: {Evolutionary
  accessibility of mutational pathways}.
\newblock {PLoS Comp. Biol.} \textbf{7}(8), e1002,134 (2011)

\bibitem{Gillespie1984}
Gillespie, J.H.: {Molecular evolution over the mutational landscape}.
\newblock {Evolution} \textbf{38}, 1116--1129 (1984)

\bibitem{Hegarty2014}
Hegarty, P., Martinsson, A.: {On the existence of accessible paths in various
  models of fitness landscapes}.
\newblock {Ann. Appl. Probab.} \textbf{24}, 1375--1395 (2014)

\bibitem{Hwang2018} Hwang, S., Schmiegelt, B., Ferretti, L., Krug, J.:
  {Universality classes of interaction structures for NK fitness landscapes}.
  \newblock {J. Stat. Phys.} \textbf{172}, 226--278 (2018) 

\bibitem{Kauffman1987}
Kauffman, S., Levin, S.: {Towards a general theory of adaptive walks on rugged
  landscapes}.
\newblock Journal of Theoretical Biology \textbf{128}(1), 11--45 (1987)

\bibitem{Kingman1978}
Kingman, J.F.C.: {A simple model for the balance between selection and
  mutation}.
\newblock Journal of Applied Probability \textbf{15}(1), 1--12 (1978)

\bibitem{Kistler2020}
Kistler, N., Schertzer, A.: {Undirected polymers in random environment: Path properties in the mean field limit}.
\newblock Preprint arXiv:2012.04076 (2020)

\bibitem{Krug2019}
  Krug, J.: {Accessibility percolation in random fitness landscapes}.
  \newblock In: {Probabilistic Structures in Evolution}, ed. by E. Baake and A. Wakolbinger (EMS Press, 2021)

\bibitem{Li2018}
Li, L.: {Phase transition for accessibility percolation on hypercubes}.
\newblock J. Theor. Prob. \textbf{31}, 2072--2111 (2018)

\bibitem{Lockhart1994}
	Lockhart, P.J., Steel, M.A., Hendy, M.D., Penny, D.: {Recovering Evolutionary Trees under a More Realistic Model of Sequence Evolution}.
	\newblock Mol. Biol. Evol. \textbf{11}, 605--612 (1994)

\bibitem{Martinsson2015}
Martinsson, A.: {Accessibility percolation and first-passage site percolation
  on the unoriented binary hypercube}.
\newblock Preprint arXiv:1501.02206 (2015)

\bibitem{Martinsson2016}
Martinsson, A.: {Unoriented first-passage percolation on the n-cube}.
\newblock Ann. Prob. \textbf{26}, 2597--2625 (2016)

\bibitem{Martinsson2018}
Martinsson, A.: {First-passage percolation on Cartesian power graphs}.
\newblock Ann. Prob. \textbf{46}, 1004--1041 (2018)

\bibitem{Nowak2013}
Nowak, S., Krug, J.: {Accessibility percolation on $n$-trees}.
\newblock {Europhys. Lett.} \textbf{101}, 66,004 (2013)

\bibitem{Orr2002}
Orr, H.A.: {The population genetics of adaptation: the adaptation of {DNA} sequences}.
\newblock {Evolution} \textbf{56}, 1317--1330 (2002)

\bibitem{Schmiegelt2014}
  Schmiegelt, B., Krug, J.: {Evolutionary accessibility of modular fitness landscapes}.
  \newblock {J. Stat. Phys.} \textbf{154}, 334--355 (2014)

\bibitem{Szendro2013} Szendro, I.G., Schenk, M.F., Franke, J., Krug, J., de Visser, J.A.G.M.:
  {Quantitative analyses of empirical fitness landscapes}.
  \newblock{Journal of Statistical Mechanics: Theory and Experiment} \textbf{2013}, P01005 (2013)

\bibitem{deVisser2014}
de~Visser, J.A.G.M., Krug, J.: {Empirical fitness landscapes and the
  predictability of evolution}.
\newblock {Nature Reviews Genetics} \textbf{15}, 480--490 (2014)

\bibitem{Weinreich2005}
Weinreich, D.M., Watson, R.A., Chao, L.: {Sign epistasis and genetic constraint
  on evolutionary trajectories}.
\newblock {Evolution} \textbf{59}, 1165--1174 (2005)

\bibitem{Wu2016}
Wu, N.C., Dai, L., Olson, C.A., Lloyd-Smith, J.O., Sun, R.: {Adaptation in
  protein fitness landscapes is facilitated by indirect paths}.
\newblock eLife \textbf{5}, 16,965 (2016)

\bibitem{Zagorski2016}
Zagorski, M., Burda, Z., Waclaw, B.: {Beyond the hypercube: evolutionary
  accessibility of fitness landscapes with realistic mutational networks}.
\newblock {PLoS Comp. Biol.} \textbf{12}(12), e1005,218 (2016)

\end{thebibliography}
\end{document}